\documentclass[aps,prx,reprint,superscriptaddress,nofootinbib,floatfix]{revtex4-2}
\ifdefined\pdfminorversion\pdfminorversion=7\fi
\usepackage{amsmath,amssymb,bm}
\usepackage{graphicx}
\usepackage{booktabs}
\usepackage{longtable}
\usepackage{array}
\usepackage{xcolor}
\usepackage{placeins}
\usepackage{CJKutf8}
\usepackage{microtype}
\usepackage[colorlinks=true,linkcolor=blue,citecolor=blue,urlcolor=blue,hypertexnames=false]{hyperref}
\usepackage{physics}
\usepackage{cleveref}

\crefname{equation}{Eq.}{Eqs.}
\crefname{figure}{Fig.}{Figs.}
\crefname{section}{Sec.}{Sec.}
\begin{document}
\raggedbottom

\newcommand{\jeet}[1]{\textcolor{blue}{Jeet: #1}}
\newcommand{\CF}[1]{\textcolor{red}{Chris: #1}}
\newcommand{\peter}[1]{\textcolor{purple}{Peter: #1}}
\newcommand{\JL}[1]{\textcolor{teal}{Jacob: #1}}
\newcommand{\YX}[1]{\textcolor{orange!85!black}{Yuxin: #1}}
\newcommand{\AVG}[1]{\textcolor{magenta}{Alexey: #1}}
\newcommand{\ZYW}[1]{\textcolor{green!50!black}{ZYW: #1}}

\newcommand{\papertitle}{Strong-to-Weak Spontaneous Symmetry Breaking Transitions on a Quantum Processor}
\title{\texorpdfstring{\resizebox{\textwidth}{!}{\papertitle}}{\papertitle}}

\author{Zhi-Yuan Wei\begin{CJK}{UTF8}{gbsn} (魏志远)\end{CJK}}
  \email[Corresponding author: ]{zywei@umd.edu}
    \affiliation{Joint Quantum Institute, NIST/University of Maryland, College Park, MD, 20742, USA}\affiliation{Joint Center for Quantum Information and Computer Science, NIST/University of Maryland, College Park, MD, 20742, USA}

\author{Christopher Fechisin}
  \affiliation{Joint Quantum Institute, NIST/University of Maryland, College Park, MD, 20742, USA}\affiliation{Joint Center for Quantum Information and Computer Science, NIST/University of Maryland, College Park, MD, 20742, USA}
  
\author{Yu-Xin Wang \begin{CJK}{UTF8}{gbsn} (王语馨)\end{CJK}}
  \affiliation{Joint Center for Quantum Information and Computer Science, NIST/University of Maryland, College Park, MD, 20742, USA}
  
\author{Jeet Shah}
\affiliation{Joint Quantum Institute, NIST/University of Maryland, College Park, MD, 20742, USA}\affiliation{Joint Center for Quantum Information and Computer Science, NIST/University of Maryland, College Park, MD, 20742, USA}

    \author{Tsung-Cheng Lu}
  \affiliation{Joint Center for Quantum Information and Computer Science, NIST/University of Maryland, College Park, MD, 20742, USA}
  \affiliation{Department of Physics and Astronomy, Center for Quantum Research and Technology,
University of Oklahoma, Norman, OK 73069, USA}

  \author{Cheng-Ju Lin}
  \affiliation{Joint Quantum Institute, NIST/University of Maryland, College Park, MD, 20742, USA}\affiliation{Joint Center for Quantum Information and Computer Science, NIST/University of Maryland, College Park, MD, 20742, USA}
  \affiliation{BlueQubit Inc., San Francisco, CA, 94105, USA}
  \affiliation{Department of Physics, University of Miami, Coral Gables, FL, 33146, USA}

\author{Alexey~V.~Gorshkov}
  \affiliation{Joint Quantum Institute, NIST/University of Maryland, College Park, MD, 20742, USA}\affiliation{Joint Center for Quantum Information and Computer Science, NIST/University of Maryland, College Park, MD, 20742, USA}

\begin{abstract}
Mixed quantum states can exhibit unconventional forms of symmetry breaking without pure-state counterparts. A prominent example is strong-to-weak spontaneous symmetry breaking (SW-SSB), which cannot be diagnosed by observables linear in the density matrix. Here, we experimentally resolve finite-size critical signatures of SW-SSB transitions on a superconducting quantum processor. We develop a measurement protocol that combines quantum measurements with classical processing based on the known dephasing channel to reconstruct nonlinear mixed-state correlators. Using only two measurement settings,
the protocol reconstructs fidelity,
R\'enyi-2, and linear correlators,
as well as string observables. Applying this approach to matrix-product states
subjected to power-law-correlated dephasing,
we observe finite-size crossings and scaling
collapses in nonlinear correlators, accompanied
by decaying linear correlations.
We further map the estimated SW-SSB transition
points across initial-state and channel parameters.
Measurements of linear and R\'enyi-2 string observables further reveal signatures of mixed-state symmetry-protected topological (SPT) phases, with a nonlinear string-order crossover aligned with an independently estimated SW-SSB transition point. Our results establish a quantum-classical approach to probing SW-SSB critical behavior and mixed-state SPT order.
\end{abstract}

\maketitle
\section{Introduction}
Spontaneous symmetry breaking is a central principle for organizing quantum phases of matter. Extending this principle to mixed states reveals new forms of order that lie beyond conventional descriptions of pure-state phases. A prominent example is strong-to-weak spontaneous symmetry breaking (SW-SSB), in which a strong symmetry is spontaneously broken while the corresponding weak symmetry remains intact~\cite{lessa2025swssb,sala2024purification}. Unlike conventional symmetry breaking, SW-SSB cannot be diagnosed by quantities linear in the density matrix but instead requires quantities that depend nonlinearly on it.  SW-SSB has also been connected to information-theoretic transitions in quantum memories and emergent hydrodynamics in decohered systems~\cite{lee2023quantum,sala2024purification,bao2023mixed-state,lessa2025swssb,huang2025hydrodynamics,vijay2025informationcriticalphasesdecoherence,hauser2026strongtoweaksymmetrybreakingopen}.

Experimentally, SW-SSB has recently been observed in a dephased Fermi gas~\cite{wang2026fermigas}, while other mixed-state phases have been probed on quantum processors using R\'enyi correlators and variational decoding~\cite{zhang2026probing}. However, experimentally characterizing the critical behavior of SW-SSB transitions directly through nonlinear correlators remains an outstanding challenge. Such characterization requires measurements over a range of
noise strengths and system sizes, while general hardness
results constrain SW-SSB detection without prior
knowledge of the state~\cite{feng2025hardness}.

\begin{figure}[!tbp]
 \includegraphics[width=\columnwidth]{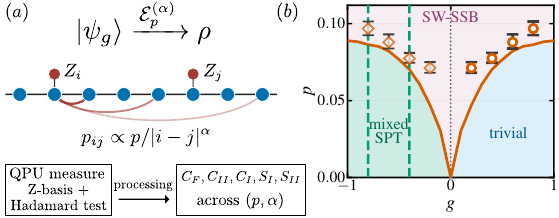}
 \caption{\label{fig:overview}
\textbf{Setup and phase diagram.}
(a) We study mixed states $\rho$ generated by applying a power-law-correlated dephasing channel $\mathcal{E}_p^{(\alpha)}$ [\cref{eq:channel}] to initial matrix-product states (MPS) $|\psi_g\rangle$ [\cref{eq:mps}]. Computational-basis measurements provide
the data for reconstructing the SW-SSB
correlators $C_F$, $C_{\mathrm{II}}$,
and $C_{\mathrm I}$, while an additional
Hadamard-test measurement provides access
to the string observables $S_{\mathrm I}$
and $S_{\mathrm{II}}$. The reconstruction combines quantum measurements with classically computed channel weights, allowing different $(p,\alpha)$ to be explored using the same measurement data.
(b) Phase diagram at $\alpha=1.4$. The orange curve indicates the theoretical thermodynamic-limit SW-SSB transition boundary obtained from the R\'enyi-2 correlator, and the shaded regions denote the corresponding mixed-state phases~\cite{fechisin2026powerlaw}. Open circles indicate transition points estimated from quantum processing unit (QPU) R\'enyi-2 correlators; diamonds show their counterparts under the model's $g\leftrightarrow-g$ duality. Green cuts at $g=-0.4$ and $-0.8$ indicate the parameters used for measurements of string observables probing mixed-SPT signatures.}
\end{figure}

In our previous work~\cite{fechisin2026powerlaw}, we introduced a tunable setting for SW-SSB transitions based on matrix-product states (MPS) subjected to power-law-correlated dephasing [Fig.~\ref{fig:overview}(a)]. The model exhibits finite-noise SW-SSB transitions and provides a three-parameter space $(g,p,\alpha)$ for systematically investigating phase boundaries and critical behavior. The same model also hosts a mixed-state
symmetry-protected topological (SPT) regime
~\cite{degroot2022openspt,ma2025mixedspt,
zhang2025communication,fechisin2026powerlaw}
[Fig.~\ref{fig:overview}(b)]. Despite these theoretical predictions,
experimentally accessing this phase structure
through nonlinear mixed-state observables
calls for efficient measurement protocols
across system sizes and channel parameters.

Here, we report the first experimental observation of finite-size critical scaling at a finite-noise SW-SSB transition on a superconducting quantum processor, using systems of up to 12 qubits. We develop a quantum-classical measurement
protocol that separates channel-independent
quantum measurements from classical processing
of the dephasing channel. The measured data can be reused across channel parameters without repeated channel sampling: a single computational-basis measurement setting provides access to fidelity, R\'enyi-2, and linear correlators, while one additional Hadamard-test setting enables measurements of string observables. Using this approach, we experimentally
characterize finite-size SW-SSB critical
behavior and map the corresponding
transition points across a tunable
parameter space.
Complementary string-order measurements
provide evidence for mixed-state SPT
order and its connection to SW-SSB.

\section{Mixed-state model and order diagnostics}

We study an open chain of $N$ qubits initialized
in a bond-dimension-two matrix-product state
(MPS)~\cite{wolf2006mps}:
\begin{equation}
 |\psi_g\rangle=\frac{1}{\sqrt{\mathcal N_g}}
 \sum_{\boldsymbol\sigma\in\{0,1\}^N}
 \langle L|M_g^{[\sigma_1]}\cdots M_g^{[\sigma_N]}|R\rangle
 |\boldsymbol\sigma\rangle,
 \label{eq:mps}
\end{equation}
with tensors
\begin{equation}
 M_g^{[0]}=\begin{pmatrix}0&0\\1&1\end{pmatrix},\qquad
 M_g^{[1]}=\begin{pmatrix}1&g\\0&0\end{pmatrix}.
 \label{eq:mps-tensors}
\end{equation}
Here, \(\mathcal N_g\) is the normalization factor, and the boundary vectors are \(\langle L|=(g^{-1/2},1)\), \(|R\rangle=(\sqrt g,1)^T\). For \(g<0\), we take \(\sqrt g=i\sqrt{|g|}\) and define the \(g=0\) state by continuity; see the Supplemental Material (SM)~\cite{supp} for details. The family interpolates between the cluster state ($g=-1$), the GHZ state ($g=0$), and the product state $|+\rangle^{\otimes N}$ ($g=1$). Denoting the Pauli operators at site $i$  by $X_i,Y_i,Z_i$, 
the states are invariant under the global
spin-flip symmetry $U_X=\prod_i X_i$.
The bulk SPT phase for $g<0$ is
protected by the antiunitary symmetry
$\mathcal T=U_X\mathcal K$, where
$\mathcal K$ denotes complex conjugation~\cite{smith2022crossing,fechisin2026powerlaw} in the computational basis. These states can be efficiently prepared using several approaches~\cite{scheer2025renormalization,Schon2005,wei2025state,malz2023preparation,smith2024constant,zyadi,tang2026weighted}. In this work, we prepare them using a sequential quantum circuit~\cite{Schon2005,supp}.

We investigate the mixed state $\rho=\mathcal{E}_p^{(\alpha)}(|\psi_g\rangle\langle\psi_g|)$ generated by power-law-correlated dephasing,
\begin{equation}
 \mathcal{E}_p^{(\alpha)}=\prod_{i<j}\mathcal E_{ij},\quad
 \mathcal E_{ij}(\rho)=(1-p_{ij})\rho+p_{ij}Z_iZ_j\rho Z_iZ_j,
 \label{eq:channel}
\end{equation}
with $p_{ij}=p/[K_N(\alpha)|i-j|^\alpha]$,
where $K_N(\alpha)=N^{1-\alpha}$ for
$0\leq\alpha\leq1$ and $K_N(\alpha)=1$
for $\alpha>1$, following
Ref.~\cite{fechisin2026powerlaw}.

{
The channel has a strong global spin-flip symmetry:
its Kraus operators commute with
$U_X=\prod_i X_i$.
Since the initial MPS satisfies
$U_X|\psi_g\rangle=|\psi_g\rangle$,
the dephased state obeys
$U_X\rho=\rho$, and hence also
$U_X\rho U_X^\dagger=\rho$.
}
For $g\neq0$, our theory predicts
SW-SSB transitions over a broad range
of power-law exponents, as diagnosed
by R\'enyi-2 correlators. For $g<0$, the low-noise regime exhibits
mixed-state SPT order, characterized by
R\'enyi-2 string correlations
~\cite{fechisin2026powerlaw}
[Fig.~\ref{fig:overview}(b)].
We use this tunable family to probe finite-size
signatures of SW-SSB critical behavior and
mixed-state SPT order.

We characterize SW-SSB using two nonlinear
correlation functions: the fidelity correlator~\cite{lessa2025swssb},
\begin{equation}
 C_F(i,j)=\mathrm{Tr}\sqrt{\sqrt\rho\,O\rho O^\dagger\,\sqrt\rho} = \|\sqrt{\rho}O\sqrt{\rho}\|_1~,
 \label{eq:fidelity}
\end{equation}
where $O=Z_iZ_j$ and $\|X\|_1 = \mathrm{Tr}\sqrt{X^{\dagger}X}$,
and the R\'enyi-2 correlator
~\cite{sala2024purification,lessa2025swssb,sun2025scheme,fechisin2026powerlaw},
\begin{align}
 C_{\mathrm{II}}(i,j)=\frac{\mathrm{Tr}(\rho O\rho O^\dagger)}{\mathrm{Tr}(\rho^2)}.
 \label{eq:correlators}
\end{align}
For comparison, we also measure the linear correlator
$C_{\mathrm I}(i,j)=\mathrm{Tr}(\rho O)$.

The fidelity correlator $C_F$ quantifies the
indistinguishability of $\rho$ and the transformed
state $O\rho O^\dagger$.
Here, $O=Z_i Z_j$ consists of two spatially separated
local operators, each charged under the strong symmetry, but not the weak.
SW-SSB is defined by a nonzero value of $C_F$
at long distances along with a vanishing linear correlator
$C_{\mathrm I}$.
The R\'enyi-2 correlator $C_{\mathrm{II}}$
probes symmetry breaking in the
doubled-Hilbert-space representation of the
mixed state.
Although the two diagnostics
are related, they need not share the same transition point or
critical behavior. 
In our setting, $C_{\mathrm{II}}$ admits a mapping to a classical
statistical-mechanics model with long-range
Ising interactions, yielding predictions
for the phase boundary and critical
scaling~\cite{fechisin2026powerlaw}.

To probe mixed-state SPT order, we use the
string operator~\cite{smith2022crossing,fechisin2026powerlaw}
\begin{equation}
 \mathcal S_{mn}=Z_{m-1}Y_m\!\left(\prod_{k=m+1}^{n-1}X_k\right)Y_nZ_{n+1},
 \label{eq:string}
\end{equation}
and define the linear and R\'enyi-2 string order parameters as
\begin{equation}
 S_{\mathrm{I}}=\mathrm{Tr}(\rho\mathcal S_{mn}),\qquad
 S_{\mathrm{II}}=\frac{\mathrm{Tr}(\mathcal S_{mn}\rho^2)}
 {\mathrm{Tr}(\rho^2)} .
 \label{eq:stringorders}
\end{equation}
For the family of initial states and channels considered here,
our theory predicts a regime in which the R\'enyi-2 string order remains nonzero in the
long-string limit, while the corresponding
linear string order vanishes $S_{\mathrm I}\to0$~\cite{fechisin2026powerlaw}.

\section{Measuring nonlinear mixed-state observables}

{
Experimentally mapping SW-SSB transitions
requires nonlinear measurements across
system sizes and noise strengths.
We develop a quantum-classical measurement
protocol that combines channel-independent
quantum measurements with efficient
classical processing of the dephasing
channel. This allows mixed-state observables
to be reconstructed throughout the $(p,\alpha)$
parameter space from the same QPU data at
fixed $g$, avoiding repeated channel sampling
and its associated fluctuations.
}

{
The dephasing channel admits a compact
representation in terms of Pauli error
patterns. Each pairwise channel
$\mathcal E_{ij}$ in Eq.~\eqref{eq:channel}
independently applies either $I$ or
$Z_iZ_j$ with probabilities $1-p_{ij}$
and $p_{ij}$, respectively.
Since these Pauli errors commute and
$Z_i^2=I$, any channel realization is
equivalent to a net error
$Z(\mathbf{s})=\prod_i Z_i^{s_i}$,
specified by a bitstring
$\mathbf{s}=(s_1,\ldots,s_N)$.
Each pairwise error changes two entries
of $\mathbf s$, so only even-parity
bitstrings can occur.
}
The resulting mixed state can be written as
\begin{equation}
 \rho=\sum_{\mathbf{s}:\,|\mathbf{s}|\ {\rm even}}
 q_{p,\alpha}(\mathbf{s})
 Z(\mathbf{s})|\psi_g\rangle
 \langle\psi_g|Z(\mathbf{s}),
 \label{eq:rho-mask}
\end{equation}
where $q_{p,\alpha}(\mathbf{s})$ is the
probability of the net Pauli error pattern
$Z(\mathbf{s})$. These probabilities depend
only on $(p,\alpha)$ and can be computed
classically~\cite{supp}.

{
The Walsh--Hadamard transform of the
channel-error distribution has a compact
product form,
\begin{equation}
\begin{aligned}
{\cal K}(\mathbf{w})
=\sum_{\mathbf{s}:\,|\mathbf{s}|\ {\rm even}}
q(\mathbf{s})(-1)^{\mathbf{s}\cdot\mathbf{w}}
=\prod_{k<l:\,w_k\neq w_l}(1-2p_{kl}),
\end{aligned}
\label{eq:channel-kernel}
\end{equation}
Here, $\mathbf w$ is a binary bitstring,
the dependence of $q(\mathbf{s})$ on $(p,\alpha)$
is implicit, and
$\mathbf s\cdot\mathbf w$ denotes
the binary inner product.
Each independent pairwise channel contributes
a factor $1-2p_{kl}$ when $w_k\neq w_l$,
and unity otherwise.
Thus, ${\cal K}(\mathbf{w})$ can be evaluated
with $O(N^2)$ classical operations,
without enumerating the channel-error patterns
(see SM~\cite{supp}).
}

To reconstruct the fidelity correlator $C_F(i,j)$, we prepare
$|\psi_g\rangle=U_g|0\rangle^{\otimes N}$
using a sequential MPS preparation
circuit~\cite{Schon2005} and measure all qubits
in the computational basis.
From $M$ measured $N$-bit outcomes
$\mathbf{z}_1,\ldots,\mathbf{z}_M$, we construct
the $M\times M$ matrix $G$ and a diagonal matrix $D$
\begin{equation}
    G_{ab}=\frac{{\cal K}(\mathbf{z}_a\oplus \mathbf{z}_b)}{M}~, 
\qquad
D_{ab}=\delta_{ab}
(-1)^{z_{a,i}+z_{a,j}}~,
\end{equation}
where $\oplus$ denotes bitwise addition
modulo two (XOR) and $z_{a,i}$ is the bit at site $i$ in
measurement outcome $z_a$.
Here, $G$ encodes the dephasing channel through the kernel ${\cal K}$, while $D$ encodes the probe $Z_iZ_j$.
As derived in SM~\cite{supp},
we estimate the fidelity correlator
from the measured bitstrings as
\begin{equation}
\widehat C_F=
\bigl\|G^{1/2}DG^{1/2}\bigr\|_1~.
\label{eq:main-gram}
\end{equation}
This formulation avoids explicitly
constructing the exponentially large distribution
of channel-error patterns.
The reconstruction applies to arbitrary
initial pure states under the specified
dephasing channel and a probe operator
diagonal in the $Z$ basis.

{The same computational-basis measurements also suffice to reconstruct  the R\'enyi-2 and linear
correlators. 
The linear correlator is simply
$C_{\mathrm I}(i,j)=\mathbb E_{\mathbf z}[d(\mathbf z)]$, where $d(\mathbf z)=(-1)^{z_i+z_j}$. 
The R\'enyi-2 correlator can be expressed
in terms of two independent computational-basis
measurement outcomes $\mathbf z,\mathbf z'$ as
\begin{equation}
C_{\mathrm{II}}(i,j)=
\frac{
\mathbb E_{\mathbf z,\mathbf z'}
\left[d(\mathbf z)d(\mathbf z')
{\cal K}(\mathbf z\oplus\mathbf z')^2\right]
}{
\mathbb E_{\mathbf z,\mathbf z'}
\left[{\cal K}(\mathbf z\oplus\mathbf z')^2\right]
}~,
\label{eq:maskestimator}
\end{equation}
where the denominator equals $\operatorname{Tr}(\rho^2)$.
For finite-shot estimation, we average over pairs of
independent measurement shots, excluding self-pairs
to avoid bias in the quadratic estimates
(see SM~\cite{supp}).
Thus, all three correlators $C_F$, $C_{\mathrm{II}}$,
and $C_{\mathrm I}$ can be reconstructed from
a single computational-basis measurement setting,
with classical processing costs polynomial
in $N$ and the number of recorded snapshots.

\begin{figure}[!t]
 \centering
 \includegraphics[width=\columnwidth]{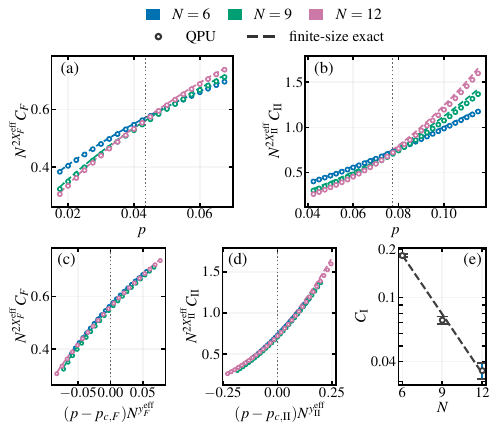}
 \caption{\label{fig:swssb}
{
\textbf{SW-SSB transition.}
Results for $g=0.4$ and $\alpha=1.4$.
(a),(b) Scaled fidelity correlator $C_F$
and R\'enyi-2 correlator $C_{\mathrm{II}}$
for $N=6,9,12$.
Colors distinguish system sizes, open symbols
denote QPU measurements, and dashed curves
show exact finite-chain calculations.
Dotted vertical lines indicate the
transition-point estimates $p_{c,F}$ and
$p_{c,\mathrm{II}}$ obtained from
finite-size scaling fits across the
three system sizes.
(c),(d) Corresponding finite-size scaling
collapses. The exact curves in (a)--(d)
are rescaled using the scaling parameters
fitted to the QPU data. Error bars in (a)–(d) are smaller
than the plotted symbols and are not shown.
(e) System-size dependence of the linear
correlator $C_{\mathrm I}$, with the dashed
curve showing exact finite-chain results.
Unscaled correlators, additional parameter
cuts, effective scaling exponents, and
fitting details are provided in the
SM~\cite{supp}.
}
}
\end{figure}

For the string order parameters $S_{\mathrm I}$ and $S_{\mathrm{II}}$, we introduce the channel-independent pure-state expectation values
\begin{align}
a_g(\mathbf u)&=\langle\psi_g|Z(\mathbf u)|\psi_g\rangle~,\\
b_g(\mathbf u)&=\langle\psi_g|\mathcal S_{mn}Z(\mathbf u)|\psi_g\rangle~,
\end{align}
where the dependence on the string endpoints $(m,n)$ is implicit in $b_g(\mathbf u)$.
Conjugation by the Pauli error $Z(\mathbf s)$ introduces a sign
$\chi(\mathbf s)=(-1)^{\sum_{j=m}^n s_j}$,
indicating whether $Z(\mathbf s)$ commutes or anticommutes
with $\mathcal S_{mn}$. The channel ensemble gives the linear
string correlator directly. For the
R\'enyi-2 string, expanding $\rho^2$
introduces two error patterns, whose
XOR difference organizes the resulting
terms~\cite{supp}:
\begin{align}
 S_{\mathrm I}&=b_g(\mathbf 0)\sum_{\mathbf s} q(\mathbf s)\chi(\mathbf s),\nonumber\\
 S_{\mathrm{II}}&=
 \frac{\sum_{\mathbf u} h(\mathbf u)a_g(\mathbf u) b_g(\mathbf u)}
 {\sum_{\mathbf u} r(\mathbf u)a_g(\mathbf u)^2}~,
 \label{eq:main-string-mask}
\end{align}
{where
$r(\mathbf u)=\sum_{\mathbf s}
q(\mathbf s)q(\mathbf s\oplus\mathbf u)$
is the probability distribution of the XOR
of two independent channel-error patterns, and
$h(\mathbf u)=\sum_{\mathbf s}
q(\mathbf s)\chi(\mathbf s)
q(\mathbf s\oplus\mathbf u)$
is its string-weighted counterpart.}

Equation~\eqref{eq:main-string-mask} shows
that the linear string correlator depends
only on $b_g(\mathbf 0)$, whereas the
nonlinear string correlator also involves
$b_g(\mathbf u)$ for nonzero
$\mathbf u$. These expectation values can be accessed through a Hadamard test.
{We prepare an ancilla in
$|+\rangle$ and apply $\mathcal S_{mn}$ to the chain conditioned on
the ancilla being in $|1\rangle$. We then measure the ancilla} in the $X$ basis, obtaining
$\sigma=\pm1$, and the chain in the
computational basis, obtaining $\mathbf z$. 
The joint measurement outcomes yield
$\mathbb E_{\sigma,\mathbf z}
\left[\sigma(-1)^{\mathbf u\cdot\mathbf z}\right]
=\operatorname{Re} b_g(\mathbf u)$.
Combining the joint Hadamard-test outcomes
with an independent set of $Z$-basis snapshots
allows us to reconstruct both string
correlators using the channel kernel
${\cal K}$ defined in
Eq.~\eqref{eq:channel-kernel}. Further details are provided in the SM~\cite{supp}.

All five observables require only two measurement
settings. For the MPS family studied here, the
preparation and controlled-string circuits have
gate complexity linear in $N$, while classical
post-processing scales polynomially with $N$
and the number of recorded snapshots.
These resource scalings do not establish a
polynomial shot complexity for a fixed
measurement precision, which also depends
on the state, channel, and observable.

\section{SW-SSB Critical scaling}
Using the protocol described above,
we acquire quantum measurement data
on superconducting quantum processors
for $g\in\{0.2,0.4,0.6,0.8\}$ and
$N\in\{6,9,12\}$.
{From the measured snapshots and the computed channel kernel, we reconstruct} $C_F(i,j)$,
$C_{\mathrm{II}}(i,j)$, and $C_{\mathrm I}(i,j)$
at $i=N/3$ and $j=2N/3$, corresponding to a separation that increases with system size.
{Details of the circuits, shot allocations, and parallel measurement
layouts are provided in the SM~\cite{supp}.}

\begin{figure}[!t]
 \centering
 \includegraphics[width=\columnwidth]{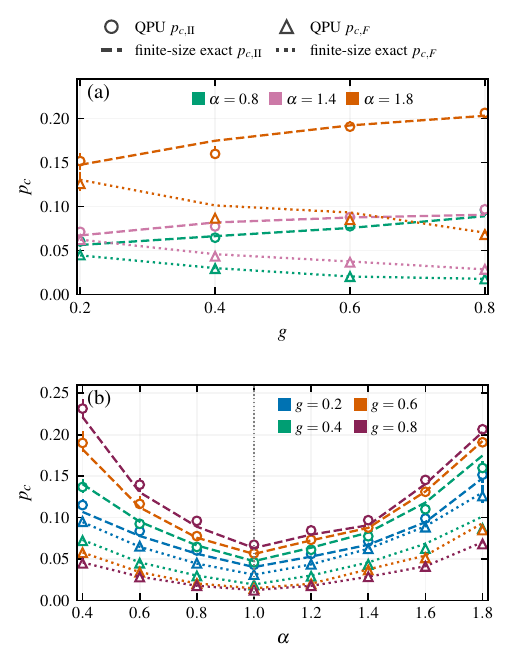}
\caption{\label{fig:phase}
{
\textbf{Estimated SW-SSB transition points.}
(a) Dependence of the estimated transition
points $p_{c,F}$ and $p_{c,\mathrm{II}}$
on the MPS parameter $g$ at fixed $\alpha$.
The $\alpha=1.4$ cut corresponds to the
phase diagram in Fig.~\ref{fig:overview}(b).
(b) Dependence on the power-law exponent
$\alpha$ at fixed $g$.
The dotted vertical line marks $\alpha=1$,
where the channel normalization changes
its functional form [cf.~\cref{eq:channel}].
Symbols denote QPU estimates, while dashed
and dotted curves show transition-point
estimates obtained from separate fits to
the exact finite-chain correlators using
the same fitting procedure.
A comparison with large-system theoretical
phase boundaries is provided in the
SM~\cite{supp}.
}
}
\end{figure}

{
Figure~\ref{fig:swssb} shows representative
results at $g=0.4$ and $\alpha=1.4$.
The scaled fidelity and R\'enyi-2 correlators
exhibit clear finite-size crossings
[Figs.~\ref{fig:swssb}(a),(b)].
Finite-size scaling fits of the three
measured sizes $N=6,9,12$ yield
transition-point estimates $p_{c,D}$
($D=F,\mathrm{II}$), with the corresponding
scaling collapses shown in
Figs.~\ref{fig:swssb}(c),(d).
In contrast, the linear correlator
$C_{\mathrm I}$ decays approximately
exponentially with $N$
[Fig.~\ref{fig:swssb}(e)].
The measurements agree well with exact
finite-chain results (dashed curves).
Additional parameter cuts and fitting
details are provided in the SM~\cite{supp}.
Together, these results reveal finite-size
critical signatures of SW-SSB transitions.
}

\begin{figure*}[!t]
 \centering
 \includegraphics[width=0.965\textwidth]{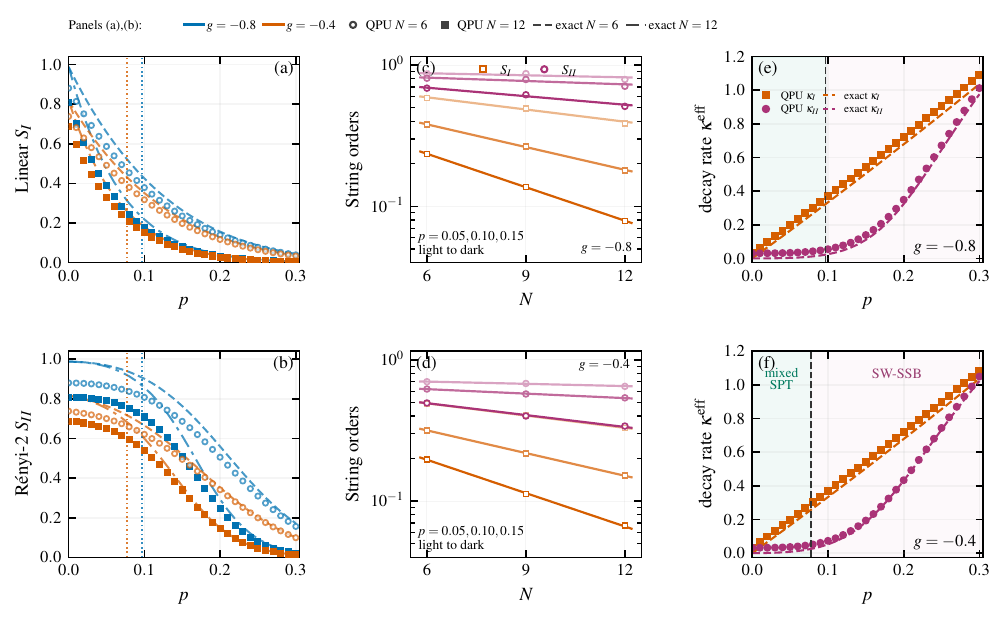}
 \caption{\label{fig:strings}
{
\textbf{Signatures of the mixed-SPT phase} at $\alpha=1.4$. Error bars are smaller than the plotted symbols and are omitted. (a),(b) Linear and R\'enyi-2 string order parameters,
$S_{\mathrm I}$ and $S_{\mathrm{II}}$, as functions of
the dephasing strength $p$ for $g=-0.8,-0.4$ and
$N=6,12$. Vertical dotted lines mark
$p_{c,\mathrm{II}}$, independently extracted from
the SW-SSB finite-size scaling at the same $|g|$
[cf.~Fig.~\ref{fig:phase}].
(c),(d) System-size dependence of the two string
orders at fixed $p=0.05,0.10,0.15$ for
$g=-0.8$ and $-0.4$, respectively.
Solid lines show exponential fits
$S_X\propto\exp[-\kappa_X^{\mathrm{eff}}\ell_N]$,
with $\ell_N=N/3$ and $X\in\{\mathrm I,\mathrm{II}\}$.
(e),(f) Corresponding effective decay rates
$\kappa_{\mathrm I}^{\mathrm{eff}}$ and
$\kappa_{\mathrm{II}}^{\mathrm{eff}}$ as functions
of $p$. Dashed curves are obtained from exact
finite-chain calculations. 
}}
\end{figure*}

We next map the estimated SW-SSB transition points
across initial-state and channel parameters.
{Using the same measured snapshots, we reconstruct
$C_F$ and $C_{\mathrm{II}}$ throughout the
$(p,\alpha)$ scan} and obtain the transition points
shown in Fig.~\ref{fig:phase}
(see the underlying data in the SM~\cite{supp}). At fixed $\alpha$, $p_{c,\mathrm{II}}$ increases overall
with $g$ [panel (a)], consistent with the
large-system theoretical trend of
Ref.~\cite{fechisin2026powerlaw}.
This trend can be understood from the
statistical-mechanics mapping: decreasing
$g$ strengthens an additional ordering
interaction, allowing SW-SSB to emerge
at weaker dephasing. We also compare the extracted
$p_{c,\mathrm{II}}$ with the thermodynamic phase boundary
in Fig.~\ref{fig:overview}(b), finding qualitative agreement despite the
modest system sizes {studied here}.
The fidelity correlator measurements reveal a distinct trend:
$p_{c,F}$ generally decreases with increasing
$g$ and lies below the corresponding R\'enyi-2 estimate,
$p_{c,F}<p_{c,\mathrm{II}}$. This finite-size ordering is consistent with the expected R\'enyi hierarchy of decoherence-induced phase transitions~\cite{yang2026mechanism,vijay2026hierarchy}.

At fixed $g$ [Fig.~\ref{fig:phase}(b)],
the estimated transition points exhibit
a nonmonotonic dependence on $\alpha$,
decreasing toward $\alpha=1$ and
increasing beyond it.
This behavior reflects the dependence
of the transition on the range and
normalization of the dephasing channel:
the normalization compensates for the
growing long-range contribution below
$\alpha=1$, while increasing $\alpha>1$
suppresses long-distance dephasing
at fixed $p$.
Moreover, $p_{c,F}$ and
$p_{c,\mathrm{II}}$ exhibit the same qualitative
dependence on $\alpha$, indicating that changing the range of the
power-law channel shifts the estimated transition points of the two nonlinear diagnostics in a similar manner.

\section{Mixed-State SPT Signatures}
We further probe mixed-state SPT order
in the same family of dephased states
using linear and R\'enyi-2 string
diagnostics [cf.~\cref{eq:stringorders}]. We prepare $|\psi_g\rangle$ at $g=-0.8,-0.4$ for $N=6,9,12$, with string endpoints $m=N/3$ and $n=2N/3$.
Figure~\ref{fig:strings} presents representative results for $\alpha=1.4$; results for other values of $\alpha$, together with further experimental details, are provided in the SM~\cite{supp}.

{Figs.~\ref{fig:strings}(a),(b) show the raw measured linear and R\'enyi-2 strings together with exact finite-chain references. Both observables decrease with increasing noise strength $p$ and system size $N$. The vertical guides mark}
$p_{c,\mathrm{II}}$, independently extracted from the
SW-SSB finite-size scaling analysis. Notably, $S_{\mathrm I}$ and $S_{\mathrm{II}}$ exhibit markedly different behavior for $p<p_{c,\mathrm{II}}$: while $S_{\mathrm I}$ decays rapidly with increasing $p$, $S_{\mathrm{II}}$ remains comparatively robust. This contrast is a characteristic signature of the mixed-SPT regime, which we quantify more systematically below through their distinct size dependences.

{Figures~\ref{fig:strings}(c),(d) show the size
dependence of $S_{\mathrm I}$ and $S_{\mathrm{II}}$
at fixed $p$, which we characterize by
$S_X\propto\exp[-\kappa_X^{\rm eff}\ell_N]$,
with $X\in\{\mathrm I,\mathrm{II}\}$ and
$\ell_N=N/3$.
The extracted decay rates $\kappa_{\mathrm I}^{\rm eff}$
and $\kappa_{\mathrm{II}}^{\rm eff}$
[Figs.~\ref{fig:strings}(e),(f)] reveal a
clear contrast: $\kappa_{\mathrm I}^{\rm eff}$
grows approximately linearly with $p$,
whereas $\kappa_{\mathrm{II}}^{\rm eff}$
remains small and weakly varying for
$p<p_{c,\mathrm{II}}$. This weak size dependence is consistent with the
long-range nonlinear string order expected
in the mixed-SPT phase.
For $p>p_{c,\mathrm{II}}$,
$\kappa_{\mathrm{II}}^{\rm eff}$ increases
more rapidly, indicating strong
suppression of nonlinear string correlations
with system size. 
The crossover in nonlinear string-order decay
aligns with the independently determined SW-SSB
transition point, providing complementary
finite-size evidence for the predicted mixed-state
phase structure and the connection between
SW-SSB and mixed-SPT order.}

{
\section{Discussion}

Our experiment resolves finite-size critical signatures
of SW-SSB transitions through two nonlinear correlators.
By systematically varying the initial-state and dephasing
parameters, we estimate the corresponding transition
points and reveal distinct behavior of fidelity and
R\'enyi-2 diagnostics.
The measurements of string correlators further provide complementary
evidence for mixed-SPT order, with a nonlinear string-order
crossover aligned with the independently extracted
SW-SSB transition in the R\'enyi-2 sense.

Beyond these observations, our quantum-classical
measurement framework offers a route toward
studying mixed-state phases in larger quantum
systems. The protocol uses only two measurement
settings, with circuit gate counts and classical
post-processing scaling polynomially with
system size and the number of recorded snapshots.
The number of shots needed to achieve
a given precision may nevertheless increase
with system size, depending on the
observable and the properties of the
mixed state, including its purity.

Our present experiments focus on $N\leq12$,
where reliable hardware measurements and
independent exact finite-chain benchmarks
are both available. Larger-system experiments
are feasible with the same protocol, but
require additional QPU resources.
Such extensions could enable more precise
characterization of SW-SSB critical behavior
and its thermodynamic scaling.
 
More broadly, the separation of quantum
measurements from channel-dependent classical
processing may be extended to other structured
noise channels and many-body systems.
Such extensions could provide new tools
for probing SW-SSB transitions and
mixed-state SPT phases in a wider range
of physical settings. They may also
enable experimental investigations of
related information-theoretic phenomena,
including decoding transitions in
quantum memories.}

\emph{Note added.---}During the preparation of this manuscript, we became
aware of two related works starting from one-dimensional
cluster states, corresponding to the $g=-1$ fixed point
of the MPS family considered here. One demonstrates
average-SPT order and finite-size SW-SSB signatures in
a five-qubit superconducting array~\cite{cao2026mixedstate}, while the other realizes
decoherence-induced average-SPT order using two-replica
string order measurements on quantum processors~\cite{shen2026averaged}.
Our experiment was developed independently of these
studies and investigates a tunable family of
non-fixed-point mixed states.
We resolve finite-size crossing and collapse
signatures of SW-SSB transitions using fidelity
and R\'enyi-2 correlators, and map their
estimated transition points.
We also measure linear and R\'enyi-2
string observables, revealing signatures of
mixed-state SPT order and a nonlinear
string-order crossover aligned with the
independently estimated SW-SSB transition.

\emph{Acknowledgments.---} Access to IBM Heron R3 quantum computers was provided by QLab, the National Quantum Laboratory and the University of Maryland.
 Z.-Y.W., C.F., J.S., and A.V.G.\ acknowledge support from the U.S.~Department of Energy, Office of Science, Accelerated Research in Quantum Computing, Fundamental Algorithmic Research toward Quantum Utility (FAR-Qu).  Z.-Y.W., C.F., J.S., and A.V.G.\ were also supported in part by the DoE ASCR Quantum Testbed Pathfinder program (award No.~DE-SC0024220), NQVL:QSTD:Design:FTL, NSF QLCI (award No.~OMA-2120757), NSF STAQ program, AFOSR MURI,  DARPA SAVaNT ADVENT, ONR MURI, and ARL (W911NF-24-2-0107). Z.-Y.W., C.F., J.S., and A.V.G.\ also acknowledge support from the U.S.~Department of Energy, Office of Science, National Quantum Information Science Research Centers, Quantum Systems Accelerator (award No.~DE-SCL0000121).  Y.-X.W.~acknowledges support from a QuICS Hartree Postdoctoral Fellowship. C.F. acknowledges the Joint Quantum Institute at the University of Maryland for support through a JQI fellowship. T.-C. L. and C.-J. L. acknowledge the support of the RQS postdoctoral fellowship through the National Science Foundation (QLCI grant OMA-2120757).

{
AI tools assisted with measurement-protocol
development, mathematical derivations,
numerical analysis, 
and manuscript preparation.
All scientific results were independently
verified by the authors, who take full
responsibility for the manuscript.
}
\bibliographystyle{apsrev4-2-publication-links}
\bibliography{references_mix_exp}

\clearpage
\onecolumngrid
\begin{center}
\textbf{Supplemental Material for ``Strong-to-Weak Spontaneous Symmetry Breaking Transitions on a Quantum Processor''}
\end{center}
\setcounter{equation}{0}
\renewcommand{\theequation}{S\arabic{equation}}
\setcounter{figure}{0}
\renewcommand{\thefigure}{S\arabic{figure}}
\setcounter{table}{0}
\renewcommand{\thetable}{S\arabic{table}}
\setcounter{section}{0}
\renewcommand{\thesection}{S\arabic{section}}
\renewcommand{\thesubsection}{\thesection.\arabic{subsection}}
\makeatletter
\renewcommand{\p@subsection}{}
\makeatother

{
This Supplemental Material provides derivations
of the measurement protocols for the five
mixed-state observables studied in the main text.
Sections~\ref{sec:S-correlator} and
\ref{sec:S-strings} derive the correlator and
string-order estimators, respectively.
Section~\ref{sec:S-implementation} details
the experimental implementation, while
Secs.~\ref{sec:S-scaling}--\ref{sec:S-alpha-strings}
present the finite-size scaling analysis,
statistical uncertainties, and additional results.
The initial MPS is defined in
Eqs.~\eqref{eq:mps} and \eqref{eq:mps-tensors},
and its phase structure under power-law
dephasing is developed in
Ref.~\cite{fechisin2026powerlaw}.
}

{\section{Kernel representation of fidelity and R\'enyi-2 correlators}
\label{sec:S-correlator}}

{We first derive the correlator identities from the channel
ensemble and then express them using a kernel evaluated directly from
measurement samples. Each derivation ends with the corresponding
finite-shot estimator; Sec.~\ref{sec:S-implementation} describes how
the samples are acquired.}

\subsection{Channel-error patterns and probabilities}

The power-law channel is built from the two-site dephasing operations
\begin{equation}
 \mathcal E_{ij}(\rho)
 =
 (1-p_{ij})\rho+p_{ij}Z_iZ_j\rho Z_iZ_j ,
 \label{eq:S-pair-channel}
\end{equation}
with
{
\begin{equation}
p_{ij}=
\begin{cases}
p/[N^{1-\alpha}|i-j|^\alpha],
&0\leq\alpha\leq1,\\
p/|i-j|^\alpha,
&\alpha>1.
\end{cases}
\label{eq:S-pij}
\end{equation}
}
This is the main-text convention
$K_N(\alpha)=N^{1-\alpha}$ for $0\leq \alpha \leq 1$ and
$K_N(\alpha)=1$ for $\alpha>1$. {
For fixed $\alpha\neq1$, the total
pair-dephasing strength scales extensively
with system size, while at $\alpha=1$
it scales as $O(N\ln N)$.
}

Each pair channel makes an independent Bernoulli choice of whether to
apply $Z_iZ_j$. Because $Z_i^2=I$, a complete channel realization is
specified only by the parity with which each site is acted on. We
represent this parity pattern by a bitstring mask
$\mathbf s=(s_1,\ldots,s_N)$ and write
\begin{equation}
 Z(\mathbf s)=\prod_i Z_i^{s_i}.
\end{equation}
Every pair error flips two mask bits, so only even-parity masks occur.
Let $q_{p,\alpha}(\mathbf s)$ denote the total probability of all pair-error
configurations that produce the same site mask $\mathbf s$. The noisy state is
then
\begin{equation}
 \rho=
 \sum_{\mathbf s:\,|\mathbf s|\ {\rm even}}
 q_{p,\alpha}(\mathbf s)
 Z(\mathbf s)|\psi_g\rangle\langle\psi_g|Z(\mathbf s).
 \label{eq:S-rho-mask}
\end{equation}
{We use this ensemble to derive the observables. The
independent pair probabilities enter the calculation through the
product kernel in Eq.~\eqref{eq:S-kernel}.}

\subsection{Fidelity from the channel ensemble and measurement samples}
\label{sec:S-fidelity}

We derive the fidelity estimator using a standard
factorization of quantum-state fidelity, closely
related to Uhlmann's theorem~\cite{watrous2018theory}.
The key simplification is that the resulting
ensemble overlaps can be evaluated directly
from computational-basis measurement samples.

For fixed channel parameters, write
$q(\mathbf s)=q_{p,\alpha}(\mathbf s)$ and
$|\psi_{\mathbf s}\rangle
=Z(\mathbf s)|\psi_g\rangle$.
The channel ensemble gives
\begin{equation}
 A=\bigl[\sqrt{q(\mathbf s)}
 |\psi_{\mathbf s}\rangle\bigr]_{\mathbf s\ {\rm even}},
 \qquad
 \rho=AA^\dagger.
 \label{eq:S-factor-A}
\end{equation}
Since $O=Z_iZ_j$ is Hermitian and unitary,
$O\rho O=(OA)(OA)^\dagger$.
The standard fidelity identity for these
factorizations yields
\begin{equation}
 C_F
 =\operatorname{Tr}
 \sqrt{\sqrt{\rho}\,O\rho O\sqrt{\rho}}
 =\|A^\dagger OA\|_1.
 \label{eq:S-fidelity-factor}
\end{equation}

To express this quantity in terms of measured
bitstrings, let
$P(\mathbf z)=|\langle\mathbf z|\psi_g\rangle|^2$
and $d(\mathbf z)=(-1)^{z_i+z_j}$.
The entries of $A^\dagger OA$ can be
written as
\begin{equation}
 A^\dagger OA
 =\sum_{\mathbf z}P(\mathbf z)d(\mathbf z)
 v_{\mathbf z}v_{\mathbf z}^\dagger,
 \qquad
 v_{\mathbf z}(\mathbf s)
 =\sqrt{q(\mathbf s)}
 (-1)^{\mathbf s\cdot\mathbf z}.
 \label{eq:S-Q-snapshot}
\end{equation}
Although these vectors are indexed by
exponentially many channel-error patterns,
only their pairwise overlaps are needed.
They are given by
\begin{equation}
\begin{aligned}
 v_{\mathbf z}^\dagger v_{\mathbf z'}
 &={\cal K}(\mathbf z\oplus\mathbf z'),\\
 {\cal K}(\mathbf w)
 &=\sum_{\mathbf s}q(\mathbf s)
   (-1)^{\mathbf s\cdot\mathbf w}\\
 &=\prod_{k<l:\,w_k\ne w_l}(1-2p_{kl}).
\end{aligned}
\label{eq:S-kernel}
\end{equation}
The product follows from the independent
pairwise dephasing errors and can be
evaluated in $O(N^2)$ operations.

Replacing $P(\mathbf z)$ by its empirical
distribution over $M$ measured bitstrings
$\mathbf z_1,\ldots,\mathbf z_M$
gives a rank-at-most-$M$ representation.
Let $V$ be the matrix whose columns are
$v_{\mathbf z_a}/\sqrt M$, and define
the Gram matrix $G=V^\dagger V$.
Since $VDV^\dagger$ and
$G^{1/2}DG^{1/2}$ have the same
nonzero eigenvalues, we obtain
\begin{equation}
 \widehat C_F
 =\|G^{1/2}DG^{1/2}\|_1,
 \qquad
 G_{ab}
 =\frac{{\cal K}(\mathbf z_a\oplus\mathbf z_b)}{M},
 \qquad
 D_{ab}=\delta_{ab}d(\mathbf z_a).
 \label{eq:S-gram-fidelity-sample}
\end{equation}
Thus, the fidelity can be reconstructed
without explicitly generating the channel-error
ensemble or forming the full density matrix.

The estimator converges to $C_F$ as the
number of independent samples increases,
although nonlinear reconstruction can
introduce finite-shot bias. The same
measurements also yield the linear correlator,
\begin{equation}
 \widehat C_{\mathrm I}
 =\frac{1}{M}\sum_{a=1}^M d(\mathbf z_a).
 \label{eq:S-CI-sample}
\end{equation}
Identical and globally complementary bitstrings
can be combined with their empirical weights,
reducing the matrix dimension to at most
$\min(M,2^{N-1})$.
Constructing the dense Gram matrix costs
$O(N^2M^2)$, followed by an $O(M^3)$
trace-norm calculation. The number of shots
needed for a given precision depends on
the state, channel, and observable.

\subsection{Scope of the fidelity correlator measurement}

{For the channel and probe considered here, the fidelity correlator
depends only on the computational-basis probabilities $P(\mathbf z)$, despite
being nonlinear in the density matrix.}

{For any initial pure state with probabilities $P(\mathbf z)$, write
\begin{equation}
 |\psi\rangle
 =
 U_\phi\sum_{\mathbf z}\sqrt{P(\mathbf z)}\,|\mathbf z\rangle ,
 \qquad
 U_\phi=\sum_{\mathbf z} e^{i\phi_{\mathbf z}}|\mathbf z\rangle\langle \mathbf z|,
\end{equation}
{where $U_\phi$ contains the computational-basis phases.}
Both the channel masks $Z(\mathbf s)$ and the probe operator $O=Z_iZ_j$ are
diagonal in the same basis and therefore commute with {$U_\phi$}. If $\rho_0$
denotes the noisy state obtained from the corresponding phase-free input
$\sum_{\mathbf z}\sqrt{P(\mathbf z)}|\mathbf z\rangle$, then
\begin{equation}
 \rho=U_\phi\rho_0U_\phi^\dagger ,
 \qquad
 O\rho O=U_\phi(O\rho_0O)U_\phi^\dagger .
\end{equation}
Because fidelity is invariant under a common unitary conjugation,
\begin{equation}
 C_F(\rho,O\rho O)
 =
 C_F(\rho_0,O\rho_0O).
\end{equation}
Hence the fidelity correlator in this setting depends only on $P(\mathbf z)$, which explains
why the $Z$-basis sample-space estimator is sufficient.}
{The same argument applies to $C_{\mathrm{II}}$ and $C_{\mathrm I}$,
since $\mathrm{Tr}(\rho O\rho O)$, $\mathrm{Tr}(\rho^2)$, and
$\mathrm{Tr}(\rho O)$ are also invariant under the common conjugation by $U_\phi$.} This simplification does not generally apply to
the string observables, which contain off-diagonal
Pauli operators and require {the additional Hadamard-test setting} described in Sec.~\ref{sec:S-strings}.
{The reconstruction assumes that the initial
state is pure. Mixed initial states can
share the same computational-basis
distribution $P(\mathbf z)$ while yielding different
nonlinear correlators. Thus, these measurements
alone do not certify the purity of the
prepared state.}

{\subsection{Difference masks and the R\'enyi-2 correlator}
\label{sec:S-renyi}}

{The R\'enyi-2 correlator $C_{\rm II}$ involves two channel realizations
with errors $Z(\mathbf s)$ and $Z(\mathbf t)$. Their relative mask is
$\mathbf u=\mathbf s\oplus \mathbf t$, distributed according to}
{
\begin{equation}
 r_{p,\alpha}(\mathbf u)=\sum_{\mathbf s} q_{p,\alpha}(\mathbf s)q_{p,\alpha}(\mathbf s\oplus \mathbf u).
 \label{eq:S-r}
\end{equation}
}
{Because both channel masks have even parity, so does
$\mathbf u$. With $q(\mathbf s)$ and $r(\mathbf u)$ denoting the distributions at fixed
$(p,\alpha)$, the purity is}
{
\begin{align}
 \operatorname{Tr}(\rho^2)
 &=\sum_{\mathbf s,\mathbf t}q(\mathbf s)q(\mathbf t)
 |\langle\psi_g|Z(\mathbf s\oplus \mathbf t)|\psi_g\rangle|^2\nonumber\\
 &=\sum_{\mathbf u\ {\rm even}}r(\mathbf u)a_g(\mathbf u)^2.
 \label{eq:S-purity}
\end{align}
}
{For $O=Z(\mathbf o)=Z_iZ_j$, the numerator is}
{
\begin{align}
 \operatorname{Tr}(\rho O\rho O)
 &=\sum_{\mathbf s,\mathbf t}q(\mathbf s)q(\mathbf t)
 |\langle\psi_g|Z(\mathbf s\oplus \mathbf t\oplus \mathbf o)|\psi_g\rangle|^2
 \nonumber\\
 &=\sum_{\mathbf u\ {\rm even}}r(\mathbf u)a_g(\mathbf u\oplus \mathbf o)^2.
 \label{eq:S-renyi-numerator}
\end{align}
}
{Therefore,}
{
\begin{equation}
 C_{\mathrm{II}}(\mathbf o;p,\alpha)=
 \frac{\sum_u r_{p,\alpha}(\mathbf u)a_g(\mathbf u\oplus \mathbf o)^2}
 {\sum_{\mathbf u} r_{p,\alpha}(\mathbf u)a_g(\mathbf u)^2}.
 \label{eq:S-renyi-estimator}
\end{equation}
}
{This expression separates the state-dependent
overlaps $a_g(\mathbf u)$ from the channel
distribution $r_{p,\alpha}(\mathbf u)$.
Its denominator is the mixed-state purity. We next express both
parts as averages over measured bitstrings.}

{\subsection{R\'enyi-2 channel-kernel identities}
\label{sec:S-gram}}

{The channel-kernel representation of
Eq.~\eqref{eq:S-kernel} also removes the sums over relative masks
from $C_{\mathrm{II}}$. The Walsh-Hadamard transform of $r$ factorizes:}
{
\begin{equation}
 \sum_u r(\mathbf u)(-1)^{\mathbf u\cdot \mathbf w}={\cal K}(\mathbf w)^2.
\end{equation}
}
{For independent computational-basis samples $\mathbf z,\mathbf z'$ drawn
from $P(\mathbf z)=|\langle \mathbf z|\psi_g\rangle|^2$, one has
$a_g(\mathbf u)^2=\mathbb E_{\mathbf z,\mathbf z'}[(-1)^{\mathbf u\cdot(\mathbf z\oplus \mathbf z')}]$.
With $d(\mathbf z)=(-1)^{z_i+z_j}$ this gives}
{
\begin{equation}
\begin{aligned}
 \operatorname{Tr}(\rho^2)
 &=\mathbb E_{\mathbf z,\mathbf z'}[{\cal K}(\mathbf z\oplus \mathbf z')^2],\\
 \operatorname{Tr}(\rho O\rho O)
 &=\mathbb E_{\mathbf z,\mathbf z'}[d(\mathbf z)d(\mathbf z'){\cal K}(\mathbf z\oplus \mathbf z')^2].
\end{aligned}
 \label{eq:S-gram-renyi}
\end{equation}
}
{For independent snapshots, replace each expectation by
the average over distinct sample pairs. In the experiment, one hardware
shot returns $R$ simultaneous chain outcomes $\mathbf z_{tk}$, where
$t=1,\ldots,T$ and $k=1,\ldots,R$. To retain possible correlations
within a shot, define the pair average
\begin{equation}
 \mathcal U[f]=\frac{1}{T(T-1)R^2}
 \sum_{t\ne t'}\sum_{k,k'}f(\mathbf z_{tk},\mathbf z_{t'k'}).
 \label{eq:S-Ustat}
\end{equation}
The implemented estimators are
\begin{equation}
 \widehat P_2=\mathcal U[{\cal K}(\mathbf z\oplus \mathbf z')^2],\qquad
 \widehat C_{\mathrm{II}}=
 \frac{\mathcal U[d(\mathbf z)d(\mathbf z'){\cal K}(\mathbf z\oplus \mathbf z')^2]}{\widehat P_2}.
 \label{eq:S-CII-sample}
\end{equation}
Here $\widehat P_2$ estimates the mixed-state
purity $\operatorname{Tr}(\rho^2)$.
We pair outcomes from different hardware shots,
excluding pairs from the same shot even when
they belong to different spatial replicas,
to avoid potential within-shot correlations.
Assuming independent hardware shots and
identical sampling distributions, the
unnormalized estimators of
$\operatorname{Tr}(\rho^2)$ and
$\operatorname{Tr}(\rho O\rho O)$ are unbiased,
although their ratio can exhibit finite-shot
bias. Direct evaluation of these pair averages
costs $O(N^2M^2)$ for $M=TR$ recorded outcomes.
}

{\section{String-order identities and Hadamard-test coherence}
\label{sec:S-strings}}

{We derive the channel dependence of the linear and
R\'enyi-2 string observables and the coherence sampled by the
Hadamard-test circuit, followed by the finite-shot estimators.
For the initial MPS $|\psi_g\rangle$, consider the string operator}
\begin{equation}
 \mathcal S_{mn}
 =
 Z_{m-1}Y_m
 \left(\prod_{j=m+1}^{n-1}X_j\right)
 Y_nZ_{n+1}.
 \label{eq:S-physical-string}
\end{equation}
We label sites from $1$ to $N$ with
$m=N/3$ and $n=2N/3$. The resulting strings for
$N=6,9,12$ are, respectively,
\begin{equation}
 Z_1Y_2X_3Y_4Z_5,\qquad
 Z_2Y_3X_4X_5Y_6Z_7,\qquad
 Z_3Y_4X_5X_6X_7Y_8Z_9.
\end{equation}
This choice centers the string within the chain, retains the two exterior
$Z$ endpoints, and increases the string separation
$\ell_N=n-m=N/3$ with system size.

\subsection{Linear string correlator}

The essential difference from the two-point correlators in
Sec.~\ref{sec:S-correlator} is that a channel mask $Z(\mathbf s)$ need not
commute with $\mathcal S_{mn}$. Let
\begin{equation}
 \mathcal B=\{m,m+1,\ldots,n\}
\end{equation}
denote the sites on which $\mathcal S_{mn}$ contains $X$ or $Y$.
Conjugating the string by a mask gives
\begin{equation}
 Z(\mathbf s)\mathcal S_{mn}Z(\mathbf s)
 =
 \chi(\mathbf s)\mathcal S_{mn},
 \qquad
 \chi(\mathbf s)=(-1)^{|\mathbf s\cap\mathcal B|},
 \label{eq:S-chi}
\end{equation}
where
\begin{equation}
 |\mathbf s\cap\mathcal B|
 =
 \sum_{i\in\mathcal B}s_i.
\end{equation}
Thus $\chi(\mathbf s)=\pm1$ records whether the channel realization
anticommutes with the string an even or odd number of times. It obeys
the character property
\begin{equation}
 \chi(\mathbf s\oplus \mathbf t)=\chi(\mathbf s)\chi(\mathbf t).
 \label{eq:S-chi-character}
\end{equation}

We denote the linear string diagnostic by
\begin{equation}
 S_{\mathrm I}
 =
 \mathrm{Tr}(\rho\mathcal S_{mn}).
\end{equation}
Defining the pure-state string expectation value
\begin{equation}
 b_g(\mathbf 0)
 =
 \langle\psi_g|\mathcal S_{mn}|\psi_g\rangle,
\end{equation}
the channel decomposition in Eq.~\eqref{eq:S-rho-mask} gives
\begin{align}
 S_{\mathrm I}
 =
 \sum_{\mathbf s} q_{p,\alpha}(\mathbf s)
 \langle\psi_g|
 Z(\mathbf s)\mathcal S_{mn}Z(\mathbf s)
 |\psi_g\rangle
=
 b_g(\mathbf 0)
 \sum_{\mathbf s} q_{p,\alpha}(\mathbf s)\chi(\mathbf s).
 \label{eq:S-linear-string}
\end{align}
{
The linear string therefore factorizes as
$S_{\mathrm I}=b_g(\mathbf 0){\cal K}(\mathbf 1_{\mathcal B})$, where $\mathbf 1_{\mathcal B}$ is the binary bitstring
with entries equal to $1$ on $\mathcal B$ and $0$ elsewhere,
combining the measured string response
with the channel kernel.
}

\subsection{R\'enyi-2 string and compatible masks}

The R\'enyi-2 string diagnostic is
\begin{equation}
 S_{\mathrm{II}}
 =
 \frac{\mathrm{Tr}(\mathcal S_{mn}\rho^2)}
 {\mathrm{Tr}(\rho^2)}.
\end{equation}
As in the derivation of the R\'enyi-2
correlator in Sec.~\ref{sec:S-renyi},
expanding $\rho^2$ introduces two
channel-error patterns $\mathbf s$
and $\mathbf t$. Their difference
$\mathbf u=\mathbf s\oplus\mathbf t$
organizes the resulting terms.

For fixed $g$ and $N$, define
\begin{equation}
 a_g(\mathbf u)
 =
 \langle\psi_g|Z(\mathbf u)|\psi_g\rangle,
 \qquad
 b_g(\mathbf u)
 =
 \langle\psi_g|\mathcal S_{mn}Z(\mathbf u)|\psi_g\rangle.
 \label{eq:S-ab}
\end{equation}
Here, $a_g(\mathbf u)$ denotes the
pure-state overlap, while
$b_g(\mathbf u)$ includes the
string operator $\mathcal S_{mn}$.

Besides the difference-mask distribution
\begin{equation}
 r_{p,\alpha}(\mathbf u)
 =
 \sum_{\mathbf s}
 q_{p,\alpha}(\mathbf s)q_{p,\alpha}(\mathbf s\oplus \mathbf u),
 \label{eq:S-string-r}
\end{equation}
{
We also introduce the signed
difference-mask weight
}
\begin{equation}
 h_{p,\alpha}(\mathbf u)
 =
 \sum_{\mathbf s}
 \chi(\mathbf s)
 q_{p,\alpha}(\mathbf s)
 q_{p,\alpha}(\mathbf s\oplus\mathbf u).
 \label{eq:S-h}
\end{equation}
The signed weight $h$ accounts for
the commutation sign associated with
each channel-error pattern.
Its Walsh--Hadamard transform yields
the kernel factors used in the estimator.

The function $h_{p,\alpha}$ obeys a useful selection rule. Changing the
summation variable $\mathbf s\rightarrow \mathbf s\oplus \mathbf u$ and using
Eq.~\eqref{eq:S-chi-character} gives
\begin{equation}
 h_{p,\alpha}(\mathbf u)
 =
 \chi(\mathbf u)h_{p,\alpha}(\mathbf u).
 \label{eq:S-h-selection}
\end{equation}
Therefore
\begin{equation}
 h_{p,\alpha}(\mathbf u)=0
 \qquad\text{for}\qquad
 \chi(\mathbf u)=-1.
\end{equation}

{The same condition identifies which terms can contribute
to the nonlinear string. Indeed,}
{
\begin{equation}
 [\mathcal S_{mn}Z(\mathbf u)]^\dagger
 =Z(\mathbf u)\mathcal S_{mn}
 =\chi(\mathbf u)\mathcal S_{mn}Z(\mathbf u).
\end{equation}
}
{
Hence $\mathcal S_{mn}Z(\mathbf u)$ is Hermitian for
$\chi(\mathbf u)=1$, while terms with
$\chi(\mathbf u)=-1$ vanish.
Thus, $b_g(\mathbf u)$ is real
for all contributing masks,
even for complex initial pure states.
}
We now derive the numerator of $S_{\mathrm{II}}$. Write
\begin{equation}
 |\psi_{\mathbf s}\rangle=Z(\mathbf s)|\psi_g\rangle,
 \qquad
 \mathbf u=\mathbf s\oplus \mathbf t.
\end{equation}
The first matrix element is simply
\begin{equation}
 \langle\psi_{\mathbf s}|\psi_{\mathbf t}\rangle
 =
 a_g(\mathbf u).
\end{equation}
For the string matrix element,
\begin{align}
 \langle\psi_{\mathbf t}|\mathcal S_{mn}|\psi_{\mathbf s}\rangle
 =
 \langle\psi_g|
 Z(\mathbf t)\mathcal S_{mn}Z(\mathbf s)
 |\psi_g\rangle
=
 \chi(\mathbf t)
 \langle\psi_g|
 \mathcal S_{mn}Z(\mathbf t\oplus \mathbf s)
 |\psi_g\rangle
=
 \chi(\mathbf t)b_g(\mathbf u).
 \label{eq:S-string-matrix-element}
\end{align}
Consequently,
\begin{align}
 \mathrm{Tr}(\mathcal S_{mn}\rho^2)
 =
 \sum_{\mathbf s,\mathbf t}
 q_{p,\alpha}(\mathbf s)q_{p,\alpha}(\mathbf t)
 a_g(\mathbf s\oplus \mathbf t)\chi(\mathbf t)b_g(\mathbf s\oplus \mathbf t)
=
 \sum_{u\ {\rm even}}
 h_{p,\alpha}(\mathbf u)a_g(\mathbf u) b_g(\mathbf u).
 \label{eq:S-string-numerator}
\end{align}
To see the second equality, fix $\mathbf u=\mathbf s\oplus \mathbf t$, so that
$\mathbf t=\mathbf s\oplus \mathbf u$. The coefficient multiplying $a_g(\mathbf u)b_g(\mathbf u)$ is then
\begin{align}
 \sum_{\mathbf s}
 \chi(\mathbf s\oplus \mathbf u)
 q_{p,\alpha}(\mathbf s)q_{p,\alpha}(\mathbf s\oplus \mathbf u)
=
 \chi(\mathbf u)h_{p,\alpha}(\mathbf u)
=
 h_{p,\alpha}(\mathbf u),
\end{align}
where the last step uses Eq.~\eqref{eq:S-h-selection}.

The purity derived in Sec.~\ref{sec:S-renyi} is
\begin{equation}
 \mathrm{Tr}(\rho^2)
 =
 \sum_{\mathbf u\ {\rm even}}
 r_{p,\alpha}(\mathbf u)a_g(\mathbf u)^2,
 \label{eq:S-string-purity}
\end{equation}
where $a_g(\mathbf u)$ is real. Therefore
\begin{equation}
 S_{\mathrm{II}}
 =
 \frac{
 \sum_{\mathbf u\ {\rm even}}
 h_{p,\alpha}(\mathbf u)a_g(\mathbf u) b_g(\mathbf u)
 }{
 \sum_{\mathbf u\ {\rm even}}
 r_{p,\alpha}(\mathbf u)a_g(\mathbf u)^2
 }.
 \label{eq:S-string-SII}
\end{equation}
At $p=0$, only the zero mask contributes, so both
$S_{\mathrm I}$ and $S_{\mathrm{II}}$ reduce to the bare string
expectation $b_g(\mathbf 0)$.

{\subsection{Hadamard-test coherence and its kernel transform}
\label{sec:S-hadamard}}

{
To connect $b_g(\mathbf u)$ to the
Hadamard-test measurements, we express
it in the computational basis. Writing
$\psi_g(\mathbf z)=\langle\mathbf z|\psi_g\rangle$,
we obtain
\begin{equation}
 \begin{aligned}
 b_g(\mathbf u)
 &=\sum_{\mathbf z}(-1)^{\mathbf u\cdot\mathbf z}
 \,\overline{c_g(\mathbf z)},\\
 c_g(\mathbf z)
 &=\psi_g^*(\mathbf z)
 (\mathcal S_{mn}\psi_g)(\mathbf z).
 \end{aligned}
 \label{eq:S-c}
\end{equation}
}
{Preparing an ancilla in $|+\rangle$ and applying
$\mathcal S_{mn}$ to the chain conditional on ancilla state $|1\rangle$ gives
$(|0\rangle|\psi_g\rangle+|1\rangle\mathcal S_{mn}|\psi_g\rangle)/\sqrt2$.
Measuring the ancilla in the $X$ basis gives $\sigma=\pm1$, while
measuring the chain in the $Z$ basis gives $\mathbf z$. Their joint law is}
{
\begin{equation}
\begin{aligned}
 P_H(\sigma,\mathbf z)
 &=\tfrac14\bigl|\psi_g(\mathbf z)+\sigma(\mathcal S_{mn}\psi_g)(\mathbf z)\bigr|^2,\\
 \mathbb E[\sigma\mathbf1\{\mathbf z\}]&=\operatorname{Re}c_g(\mathbf z).
\end{aligned}
 \label{eq:S-hadamard}
\end{equation}
}
{The complex conjugation in Eq.~\eqref{eq:S-c} follows
from the ordering $\mathcal S_{mn}Z(\mathbf u)$ in $b_g(\mathbf u)$.
Only $\operatorname{Re}c_g$ is required for the observables:
$h_{p,\alpha}(\mathbf u)$ vanishes unless $\chi(\mathbf u)=1$, in which case
$\mathcal S_{mn}Z(\mathbf u)$ is Hermitian and $b_g(\mathbf u)$ is real.
This argument does not require real input-state amplitudes or any
MPS representation.

The Walsh-Hadamard transform of $h$ follows by setting $\mathbf t=\mathbf s\oplus \mathbf u$:
\begin{align}
 \sum_{\mathbf u} h(\mathbf u)(-1)^{\mathbf u\cdot \mathbf w}
 &=\sum_{\mathbf s,\mathbf t}\chi(\mathbf s)q(\mathbf s)q(\mathbf t)(-1)^{(\mathbf s\oplus \mathbf t)\cdot \mathbf w}\nonumber\\
 &=\left[\sum_{\mathbf s} q(\mathbf s)(-1)^{\mathbf s\cdot(\mathbf w\oplus\mathbf 1_{\mathcal B})}\right]
   \left[\sum_{\mathbf t} q(\mathbf t)(-1)^{\mathbf t\cdot \mathbf w}\right].\nonumber
\end{align}
Hence}
{
\begin{equation}
 \sum_{\mathbf u} h_{p,\alpha}(\mathbf u)(-1)^{\mathbf u\cdot \mathbf w}
 ={\cal K}(\mathbf w){\cal K}(\mathbf w\oplus\mathbf 1_{\mathcal B}),
 \label{eq:S-string-kernel-identity}
\end{equation}
}
Combining this identity with the
computational-basis expressions for
$a_g$ and $b_g$ gives
\begin{equation}
\begin{aligned}
 \operatorname{Tr}(\mathcal S_{mn}\rho^2)
 &=\sum_{\mathbf z,\mathbf z'} P(\mathbf z)\operatorname{Re}c_g(\mathbf z')\\
 &\quad\times {\cal K}(\mathbf z\oplus \mathbf z'){\cal K}(\mathbf z\oplus \mathbf z'\oplus\mathbf 1_{\mathcal B})~,
\end{aligned}
 \label{eq:S-string-gram}
\end{equation}
{
where recall $\mathbf 1_{\mathcal B}$ is the bitstring with entries equal to $1$ on
$\mathcal B=\{m,\ldots,n\}$ and $0$ elsewhere.
The linear string follows from
$S_{\mathrm I}={\cal K}(\mathbf 1_{\mathcal B})\mathbb E[\sigma]$.
For $M_Z$ recorded $Z$-basis outcomes
$\{\mathbf z_a\}$ and $M_H$ joint Hadamard-test
outcomes $\{(\sigma_b,\mathbf z'_b)\}$ from an
independent acquisition, the finite-shot
estimators are
\begin{align}
 \widehat S_{\mathrm I}
 &={\cal K}(\mathbf 1_{\mathcal B})\frac1{M_H}\sum_b\sigma_b,
 \label{eq:S-SI}\\
 \widehat S_{\mathrm{II}}
 &=\frac{\sum_{a,b}\sigma_b
 {\cal K}(\mathbf w_{ab}){\cal K}(\mathbf w_{ab}\oplus\mathbf 1_{\mathcal B})}
 {M_ZM_H\widehat P_2},
 \qquad \mathbf w_{ab}=\mathbf z_a\oplus \mathbf z'_b.
 \label{eq:S-SII}
\end{align}
The purity $\widehat P_2$ is estimated from
the $Z$-basis data using
Eq.~\eqref{eq:S-CII-sample}.
Its classical processing cost scales as
$O[N^2(M_ZM_H+M_Z^2)]$.
}

\section{Experimental implementation}
\label{sec:S-implementation}

{We describe the state-preparation circuits, the two
measurement settings, and how their outcomes enter the kernel
estimators. Statistical procedures are given in Sec.~\ref{sec:S-uncertainty}.}

\subsection{Finite-chain MPS and sequential preparation}
\label{sec:S-preparation}

The finite-chain MPS and its tensors are defined in
Eqs.~\eqref{eq:mps} and \eqref{eq:mps-tensors}. For the stated boundary
vectors, we use
$\sqrt g=i\sqrt{|g|}$ for $g<0$ and take the $g=0$ limit continuously.
With this convention,
\begin{equation}
 U_X|\psi_g\rangle=|\psi_g\rangle.
\end{equation}
For $g<0$, the bulk SPT protection is antiunitary:
$\mathcal T=U_X\mathcal K$ acts at an edge
through $V_g\mathcal K$, with
\begin{equation}
 V_g=
 \begin{pmatrix}
 0 & g\\
 1 & 0
 \end{pmatrix},
 \qquad
 V_gV_g^*=gI.
\end{equation}
The sign of $g$ therefore distinguishes the two bulk classes. The
particular open-chain edge state selected by our boundary vectors need
not itself be invariant under $\mathcal T$~\cite{smith2022crossing,fechisin2026powerlaw}.

For the experimental preparation, it is useful to write the same state
in the computational basis as
\begin{equation}
 |\psi_g\rangle=
 \frac{1}{\sqrt{2(1+|g|)^{N-1}}}
 \sum_{\boldsymbol{\sigma}}
 (\sqrt g)^{D(\boldsymbol{\sigma})}
 |\boldsymbol{\sigma}\rangle,
 \qquad
 D(\boldsymbol{\sigma})
 =
 \sum_{j=2}^N
 (\sigma_{j-1}\oplus\sigma_j).
 \label{eq:S-domain-wall}
\end{equation}
{Here $D(\boldsymbol\sigma)$ counts adjacent domain walls.
With $\theta=\arctan\sqrt{|g|}$, apply $H$ to site 1, then, for each
$j=2,\ldots,N$, apply $R_y(2\theta)$ for $g>0$ or $R_x(-2\theta)$
for $g<0$ to site $j$, followed by CNOT$_{j-1\to j}$.
We use $R_\mu(\varphi)=\exp(-i\varphi\sigma_\mu/2)$.
Each step assigns relative amplitude $\sqrt g$ to a domain wall,
giving Eq.~\eqref{eq:S-domain-wall} with $N-1$ entangling gates.
This is a sequential MPS preparation~\cite{Schon2005}.}

{\subsection{$Z$-basis acquisitions for the correlators}
\label{sec:S-correlator-measurement}}

{We prepare $|\psi_g\rangle=U_g|0^N\rangle$ and measure
every system qubit in the $Z$ basis. Let $T$ denote the number of
hardware shots and $R$ the number of disjoint spatial replicas per
shot. The recorded bitstrings are $\mathbf z_{tk}$, with $t=1,\ldots,T$
and $k=1,\ldots,R$. These snapshots determine the empirical
probabilities and the parities}
{
\begin{equation}
 \widehat P(z)=\frac1{TR}\sum_{t,k}\mathbf1[\mathbf z_{tk}=\mathbf z],\qquad
 \widehat a_g(\mathbf u)=\frac1{TR}\sum_{t,k}(-1)^{\mathbf u\cdot \mathbf z_{tk}}.
 \label{eq:S-fidelity-parities}
\end{equation}
}
{With $M=TR$, these outcomes enter
Eqs.~\eqref{eq:S-gram-fidelity-sample}, \eqref{eq:S-CI-sample}, and
\eqref{eq:S-CII-sample} for $C_F$, $C_{\mathrm I}$, and
$C_{\mathrm{II}}$, respectively.
For all system sizes we take $i=N/3$ and $j=2N/3$, with sites
numbered from $1$ to $N$. Outcomes of simultaneous replicas remain
grouped in the uncertainty analysis.}

\subsection{Hadamard-test measurements}
\label{sec:S-string-measurement}
{The string estimators of Eqs.~\eqref{eq:S-SI} and
\eqref{eq:S-SII} use one $Z$-basis data set, acquired exactly as in
Sec.~\ref{sec:S-correlator-measurement}, and one Hadamard-test data set
for each $(g,N)$.

The ancilla-conditioned operation applies $\mathcal S_{mn}$ when the
ancilla is in $|1\rangle$ and the identity when it is in $|0\rangle$.
We implement it using basis rotations and a nearest-neighbor parity
circuit. On the support sites $m-1,\ldots,n+1$, apply $H$ for each $X$
factor and $S^\dagger$ followed by $H$ for each $Y$ factor, with
$S=\operatorname{diag}(1,i)$. These rotations turn the string into
$Z_{m-1}\cdots Z_{n+1}$. A ladder of CNOT gates collects this parity
onto a site $k^\star$ coupled to the ancilla. A CZ gate between that
site and the ancilla applies the conditional phase; the ladder and
basis rotations are then undone. Preparing and measuring the ancilla
with Hadamard gates realizes its $X$-basis readout.

The joint outcomes $(\sigma,z)$ enter Eqs.~\eqref{eq:S-SI} and
\eqref{eq:S-SII}. Statevector checks reproduce the full distribution
in Eq.~\eqref{eq:S-hadamard} for the MPS inputs and generic complex
pure states, including the phases of both $Y$ factors.}
\subsection{Compilation, shots, and spatial tiling}
\label{sec:S-hardware}

{All $Z$-basis measurements were acquired on
\texttt{ibm\_pittsburgh}, and the string Hadamard measurements on
\texttt{ibm\_boston}, both 156-qubit IBM Heron R3 processors.
Circuits were compiled with Qiskit~\cite{javadiabhari2024qiskit};
the Hadamard circuits used optimization level 1.

We run identical circuits on disjoint connected sets of physical
qubits, obtaining $R$ chain outcomes per hardware shot. A Hadamard
replica contains an additional ancilla connected to the string support.
Layouts are selected using the backend calibration data and held
fixed within a campaign. Representative layouts appear in
Fig.~\ref{fig:S-layout}.

Table~\ref{tab:shots} gives the shot and replica counts. For each
negative-$g$ state we combine two 2048-shot $Z$ acquisitions with
the independent 8192-shot Hadamard acquisition. Their different
replica counts are retained in the reconstruction. All outcomes
from the same hardware shot stay grouped during uncertainty
analysis. All reported QPU observables use raw outcomes without
error mitigation.}

\begin{table}[htbp]
\centering
\caption{
Hardware-shot allocations for each $(g,N)$, with $N=6,9,12$.
Shot counts are per measurement setting except for the explicitly
aggregated $g=0.2$ $Z$-basis acquisition.
{String $Z$ shots aggregate two archived 2048-shot acquisitions.
Slash-separated replica counts correspond to $N=6/9/12$.}
}
\label{tab:shots}
\begin{tabular}{llll}
\hline
Response & Settings & Hardware shots & Spatial replicas\\
\hline
{$C_F,C_{\mathrm{II}},C_{\mathrm I}$}, $g=0.2$
& One $Z$ basis & 4096 & 8\\
{$C_F,C_{\mathrm{II}},C_{\mathrm I}$}, $g=0.4,0.6,0.8$
& One $Z$ basis & 8192 & 8\\
{String $Z$ data, $g=-0.8,-0.4$}
& {One $Z$ basis} & {4096} & {16/12/8}\\
{String Hadamard test, $g=-0.8,-0.4$}
& {One Hadamard-test setting} & {8192} & {21/15/10}\\
\hline
\end{tabular}
\end{table}

\section{Finite-size scaling analysis}
\label{sec:S-scaling}

We describe the finite-size scaling procedure and
present the complete set of nonlinear correlator
fits underlying the extracted transition-point estimates.

\subsection{Scaling ansatz and representative unscaled data}
\label{sec:S-scaling-ansatz}

To extract the finite-size transition points, we use the scaling form
motivated by the R\'enyi-2 theory of
Ref.~\cite{cardy1996scaling,fechisin2026powerlaw} and apply the same ansatz separately
to the fidelity correlator:
\begin{equation}
 C_D^{(N)}(p)
 =
 N^{-2X_D^{\mathrm{eff}}}
 \mathcal F_D
 \!\left[
 (p-p_{c,D})
 N^{y_D^{\mathrm{eff}}}
 \right],
 \qquad
 D\in\{F,\mathrm{II}\}.
 \label{eq:S-collapse}
\end{equation}
Here $p_{c,D}$ is the transition point estimated from the
three available sizes, $X_D^{\mathrm{eff}}$ controls the vertical
size dependence of the correlator near the fitted crossing, and
$y_D^{\mathrm{eff}}$ controls the rescaling of the distance from the
transition. The superscript ``eff'' on $X_D^{\mathrm{eff}}$
and $y_D^{\mathrm{eff}}$ indicates that these exponents are obtained from
$N=6,9,12$ and need not coincide with their thermodynamic values.
Equation~\eqref{eq:S-collapse} implies that
$N^{2X_D^{\mathrm{eff}}}C_D^{(N)}$ crosses near
$p_{c,D}$, while plotting the same quantity against
$(p-p_{c,D})N^{y_D^{\mathrm{eff}}}$ produces the
corresponding data collapse.
{
Figure~\ref{fig:S-unscaled} shows the
corresponding unscaled correlators
for $g=0.4$ and $\alpha=1.4$.
}

\begin{figure}[!htbp]
\centering
\includegraphics[width=0.94\textwidth]{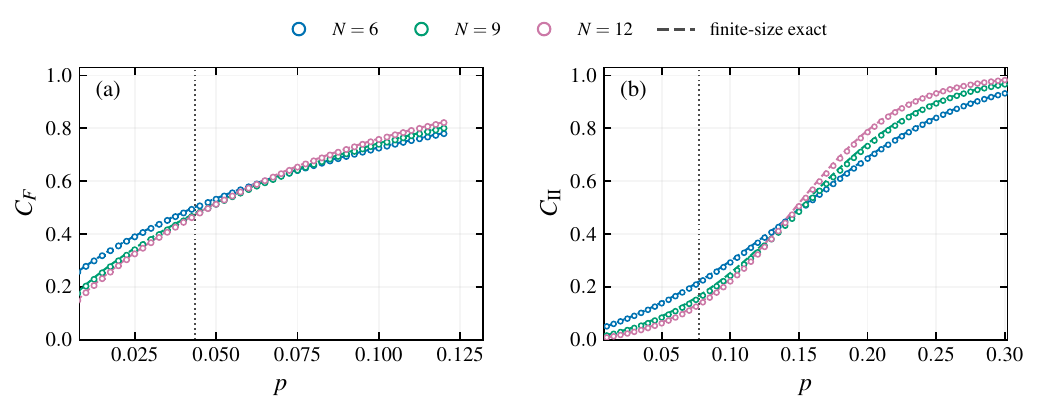}
\caption{\label{fig:S-unscaled}
Unscaled nonlinear correlators at $g=0.4$ and
$\alpha=1.4$: (a) fidelity correlator $C_F$
and (b) R\'enyi-2 correlator $C_{\mathrm{II}}$.
Open symbols denote QPU measurements for
$N=6,9,12$, and dashed curves show exact
finite-chain results.
Statistical error bars are smaller than
the symbols.
Dotted vertical lines mark the estimated
transition points obtained from the
finite-size scaling fits shown in
Fig.~\ref{fig:swssb}.
}
\end{figure}

\subsection{Implementation of the finite-size fits}
\label{sec:S-fit-implementation}

Within a fitting window of width $\Delta p$, we define
\begin{equation}
 z=
 \frac{p-p_c}{\Delta p}
 \left(\frac{N}{9}\right)^{y^{\mathrm{eff}}},
\end{equation}
and approximate the logarithm of the scaling function by
\begin{equation}
 \log C_{\mathrm{fit}}
 =
 -2X^{\mathrm{eff}}\log(N/9)
 +c_0+c_1z+c_2z^2.
 \label{eq:S-scaling-polynomial}
\end{equation}
The factors $N/9$ and $\Delta p$ set the scales
of the fit variables without changing the
ansatz in Eq.~\eqref{eq:S-collapse}.

For $C_{\mathrm{II}}$, we minimize
\begin{equation}
 J_{\mathrm{II}}
 =
 \frac{1}{M_{\mathrm{II}}}
 \sum_{N,p}
 \left[
 \frac{
 \log C_{\mathrm{II}}
 -
 \log C_{\mathrm{II,fit}}
 }{
 \sigma_{\mathrm{II}}/C_{\mathrm{II}}
 }
 \right]^2
 +
 \left[
 \frac{
 X_{\mathrm{II}}^{\mathrm{eff}}
 -
 X_{\rm ref}(\alpha)
 }{w_X}
 \right]^2,
 \qquad w_X=0.10,
 \label{eq:S-regularizer}
\end{equation}
where
\begin{equation}
 X_{\rm ref}(\alpha)
 =
 \begin{cases}
 1/4, & \alpha\leq1.5,\\
 (2-\alpha)/2, & \alpha>1.5.
 \end{cases}
\end{equation}
The reference scaling dimension is motivated
by the long-range Ising theory of
Ref.~\cite{fechisin2026powerlaw}.
Since its applicability depends on the
parameter regime, we use it as a soft
constraint, allowing
$X_{\mathrm{II}}^{\mathrm{eff}}$ to vary.
The resulting fits yield good finite-size
scaling collapses across the parameter
space [Figs.~\ref{fig:S-atlas-02-1}--
\ref{fig:S-atlas-08-2}].

For $C_F$, we use the same fitting form,
replacing the last term of
Eq.~\eqref{eq:S-regularizer} with
$[X_F^{\mathrm{eff}}/w_F]^2$, where
$w_F=0.03$. This favors weak vertical
rescaling, consistent with the observed
finite-size dependence of $C_F$.

For fixed $(p_c,y)$, the remaining
parameters are determined by constrained
linear least squares. We use multiple
initializations and impose
\begin{equation}
 0\leq X\leq0.6,\qquad
 0.1\leq y\leq2,
\end{equation}
with $p_c$ restricted to the fitting window.

The fitting windows are chosen around
the finite-size crossings, and all
transition points are extracted from
joint fits to $N=6,9,12$.
The same windows are held fixed during
statistical refits. For the exact finite-chain reference
estimates in Fig.~\ref{fig:phase},
we fit the exact correlators separately,
using the QPU-selected fitting windows
and the same weights, parameter bounds,
and regularization.
Correlations between reweighted $p$ values
are accounted for in the uncertainty analysis
of Sec.~\ref{sec:S-uncertainty}.

\subsection{Complete finite-size scaling results}
\label{sec:S-complete-scaling}

Figures~\ref{fig:S-atlas-02-1}--\ref{fig:S-atlas-08-2}
present the complete set of 32 R\'enyi-2 correlator
finite-size scaling analyses underlying
Fig.~\ref{fig:phase}, while
Figs.~\ref{fig:S-cf-atlas-02-1}--
\ref{fig:S-cf-atlas-08-2} show the
corresponding fidelity correlator results.
The dashed curves represent exact
finite-chain correlators for $N=6,9,12$,
rescaled using the scaling parameters
fitted to the QPU data.
Representative unscaled correlators are
shown in Fig.~\ref{fig:S-unscaled}.

\subsection{Uncertainty estimates}
\label{sec:S-uncertainty}

The dephasing kernel satisfies
$|{\cal K}(\mathbf w)|\leq1$, so the
pair kernels entering the unnormalized
R\'enyi-2 and string estimators are
bounded in absolute value by unity.
For independent samples, the corresponding
variances scale as $O(1/M)$ and
$O(1/M_Z+1/M_H)$, respectively.
Normalization by the mixed-state purity
can amplify statistical errors.

We estimate uncertainties by bootstrapping
the distinct-shot estimator for
$C_{\mathrm{II}}$, propagating empirical
distribution fluctuations to first order
for $C_F$, and bootstrapping the
$Z$-basis and Hadamard-test acquisitions
for the string observables.
In all cases, we treat each hardware shot,
including its spatial replicas, as the
sampling unit. Reweighting each statistical
draw over the full $p$ grid preserves
correlations between channel parameters.

Pointwise error bars denote statistical
standard errors. Fit-parameter intervals
span the 16\%--84\% of repeated
fits, with fitting windows, weights, and
regularization held fixed.

For $C_{\mathrm{II}}$, we use 180
whole-shot bootstrap draws to estimate
the statistical uncertainties of
Eq.~\eqref{eq:S-CII-sample} and the
finite-size scaling fits.

For $C_F$, we propagate fluctuations of
the empirical configuration weights
through the Gram trace norm.
With weights $w_u$ on the observed
configurations, factor
$\mathsf K_{uv}
={\cal K}(\mathbf z_u\oplus\mathbf z_v)
=(LL^T)_{uv}$ and define
\begin{equation}
 A=L^T\operatorname{diag}(w_ud_u)L,\qquad
 \widehat C_F=\|A\|_1,\qquad
 \gamma_u=d_u[L\operatorname{sign}(A)L^T]_{uu}.
 \label{eq:S-gram-influence}
\end{equation}
Here $d_u=(-1)^{z_{u,i}+z_{u,j}}$,
and $\gamma_u$ is the first-order
derivative with respect to $w_u$,
wherever the trace norm is differentiable.
Averaging the influences within each shot
gives
$v_t(p)=R^{-1}\sum_k\gamma_{z_{tk}}(p)$
and the covariance
\begin{equation}
 \widehat\Sigma_{pp'}=
 \frac{\sum_t[v_t(p)-\overline v(p)]
                  [v_t(p')-\overline v(p')]}{T(T-1)}.
 \label{eq:S-gram-covariance}
\end{equation}
We use 100 Gaussian curve draws with
this covariance to estimate the
scaling-fit parameter intervals.
The first-order approximation does not
capture the nonsmooth behavior of
the trace norm at zero eigenvalues.
For $C_{\mathrm I}$, errors are obtained
from shot-averaged parities.

For the string observables, 200 bootstrap
draws independently resample the
$Z$-basis and Hadamard-test acquisitions,
keeping spatial replicas grouped by shot.
Each draw recomputes
Eqs.~\eqref{eq:S-SI} and
\eqref{eq:S-SII}.
We extract $\kappa_X^{\mathrm{eff}}$
from weighted linear fits of
$\ln S_X$ against $\ell_N$,
and propagate the bootstrap fluctuations
to the fitted decay rates.

The reported error bars quantify statistical
sampling uncertainty and do not include
systematic hardware errors.
Throughout this work, error bars smaller
than the plotted symbols are omitted for
clarity, as noted in the corresponding
figure captions.

\section{Additional results for SW-SSB scaling}
\label{sec:S-exponents}

The effective exponents obtained from
separate fits to the QPU and exact
finite-chain correlators are compared
in Fig.~\ref{fig:S-exponents}.
Their similar trends indicate that the
measured finite-size scaling behavior
is consistent with the exact
finite-chain results.
For the R\'enyi-2 correlator,
the theoretically motivated soft constraint
allows good finite-size scaling collapses
across the measured parameter range
[Figs.~\ref{fig:S-atlas-02-1}--
\ref{fig:S-atlas-08-2}],
with fitted $X_{\mathrm{II}}^{\mathrm{eff}}$
values remaining near the reference
scaling dimensions.
At $\alpha=1.4$,
$y_{\mathrm{II}}^{\mathrm{eff}}$
decreases from approximately $1.05$
at $g=0.2$ to $0.52$ at $g=0.8$,
while the corresponding exact
finite-chain values decrease from
$1.08$ to $0.54$.
The stronger deviation from the
theoretical reference at smaller $g$
is consistent with finite-size crossover
effects.

For the fidelity correlator,
$X_F^{\mathrm{eff}}$ remains small
($\leq0.061$) under the regularization
of Sec.~\ref{sec:S-fit-implementation},
indicating weak vertical size dependence
near the fitted crossings.
The corresponding finite-size scaling
collapses are shown in
Figs.~\ref{fig:S-cf-atlas-02-1}--
\ref{fig:S-cf-atlas-08-2}.
Since no theoretical value of $X_F$
is imposed, these exponents are
interpreted as effective parameters
of the three-size scaling description
rather than asymptotic critical exponents.

Figure~\ref{fig:S-theory-alpha} places the measured $\alpha$ dependence
in the context of the large-system R\'enyi-2 phase boundaries of
Ref.~\cite{fechisin2026powerlaw}. Both show a decrease toward
$\alpha\simeq1$ followed by an increase, while the finite-chain QPU
and large-system theoretical transition points differ quantitatively.

\section{Additional data for mixed-SPT strings}
\label{sec:S-alpha-strings}

Figure~\ref{fig:S-alpha-strings} extends the
mixed-SPT string order measurements of
Fig.~\ref{fig:strings} to
$\alpha=0.8,1.4,1.8$ for $g=-0.8,-0.4$,
over $0\leq p\leq0.30$.
The $\alpha=1.4$ column corresponds to
the parameter cut studied in the main text [cf.~\cref{fig:strings}].

Across all six $(g,\alpha)$ cuts, the
linear string $S_{\mathrm I}$ is suppressed
more rapidly with increasing system size
than the R\'enyi-2 string $S_{\mathrm{II}}$
in the low-noise regime.
The dependence on $\alpha$ is also evident:
the string-order crossover occurs at
substantially larger $p$ for $\alpha=1.8$
than for $\alpha=0.8$ or $1.4$,
consistent with the corresponding shift
of the independently estimated
SW-SSB transition points.
For each $\alpha$, the crossover also
occurs at larger $p$ for $g=-0.8$
than for $g=-0.4$.

To quantify the system-size dependence,
we fit the string observables at each $p$ to
\begin{equation}
 S_X(N,p)
 \propto
 \exp[-\kappa_X^{\mathrm{eff}}(p)\ell_N],
 \qquad
 X\in\{\mathrm I,\mathrm{II}\},
 \qquad
 \ell_N=n-m=N/3.
 \label{eq:S-kappa}
\end{equation}
The three available separations
$\ell_N=2,3,4$ define effective
finite-size decay rates rather than
asymptotic exponents; in particular,
the linear string can exhibit
stretched-exponential decay for
$1<\alpha<2$~\cite{fechisin2026powerlaw}.

The extracted rates show that
$\kappa_{\mathrm{II}}^{\mathrm{eff}}$
remains comparatively small below
the independently estimated
$p_{c,\mathrm{II}}$, while
$\kappa_{\mathrm I}^{\mathrm{eff}}$
increases more steadily with $p$.
Above this boundary,
$\kappa_{\mathrm{II}}^{\mathrm{eff}}$
grows more rapidly, indicating
stronger suppression of nonlinear
string correlations.
The similar parameter dependence of
the string-order crossover and the
independently estimated SW-SSB boundary
provides further finite-size evidence
for the connection between SW-SSB
and mixed-state SPT order.

\clearpage
\newcommand{\boundaryatlas}[4]{%
\begin{figure}[p]
\centering
\includegraphics[width=0.96\textwidth,height=0.86\textheight,keepaspectratio]{figures/#1}
\caption{\label{#2}
$C_{\mathrm{II}}$ finite-size scaling at $g=#3$.
Left: scaled QPU data (symbols) and exact finite-chain calculations
(dashed) in each fitting window.
Right: the same data under the joint
$(p_{c,\mathrm{II}},
X_{\mathrm{II}}^{\mathrm{eff}},
y_{\mathrm{II}}^{\mathrm{eff}})$
rescaling, with both fitted exponents printed.
The QPU and exact results use $N=6,9,12$ and the fitting prescription
described in Sec.~\ref{sec:S-fit-implementation}.}
\end{figure}}

\boundaryatlas{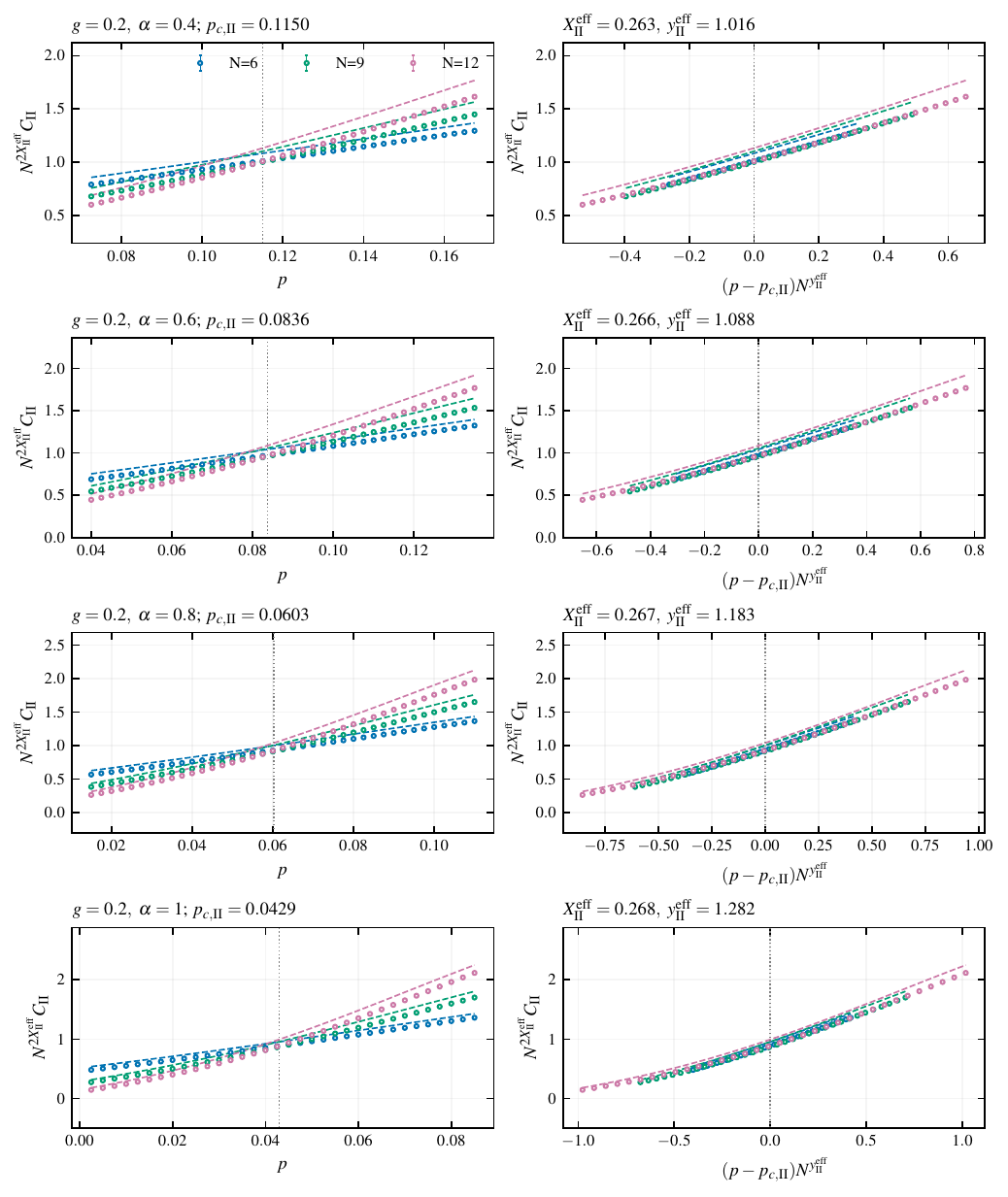}{fig:S-atlas-02-1}{0.2}{0.4,0.6,0.8,1.0}
\boundaryatlas{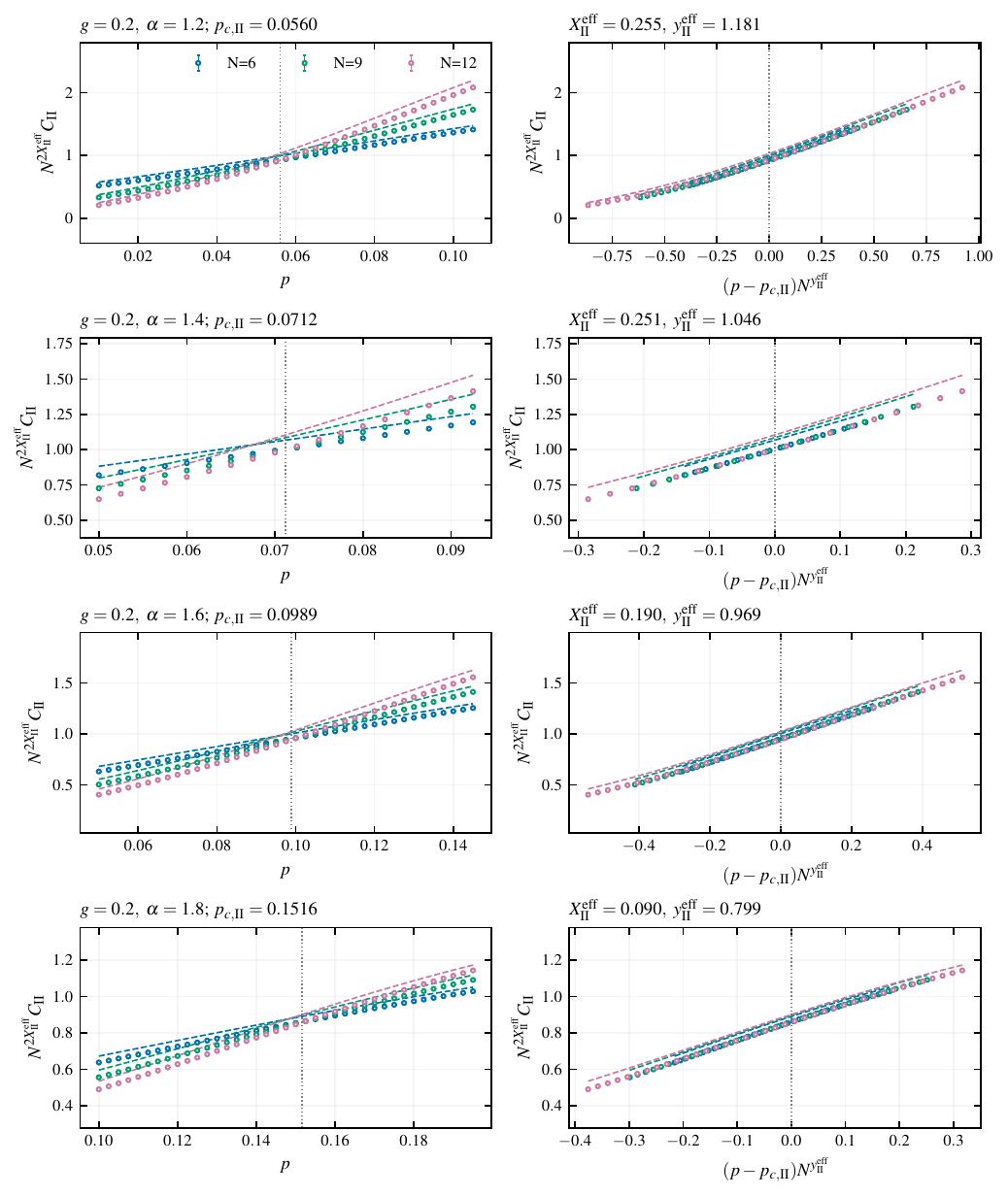}{fig:S-atlas-02-2}{0.2}{1.2,1.4,1.6,1.8}
\boundaryatlas{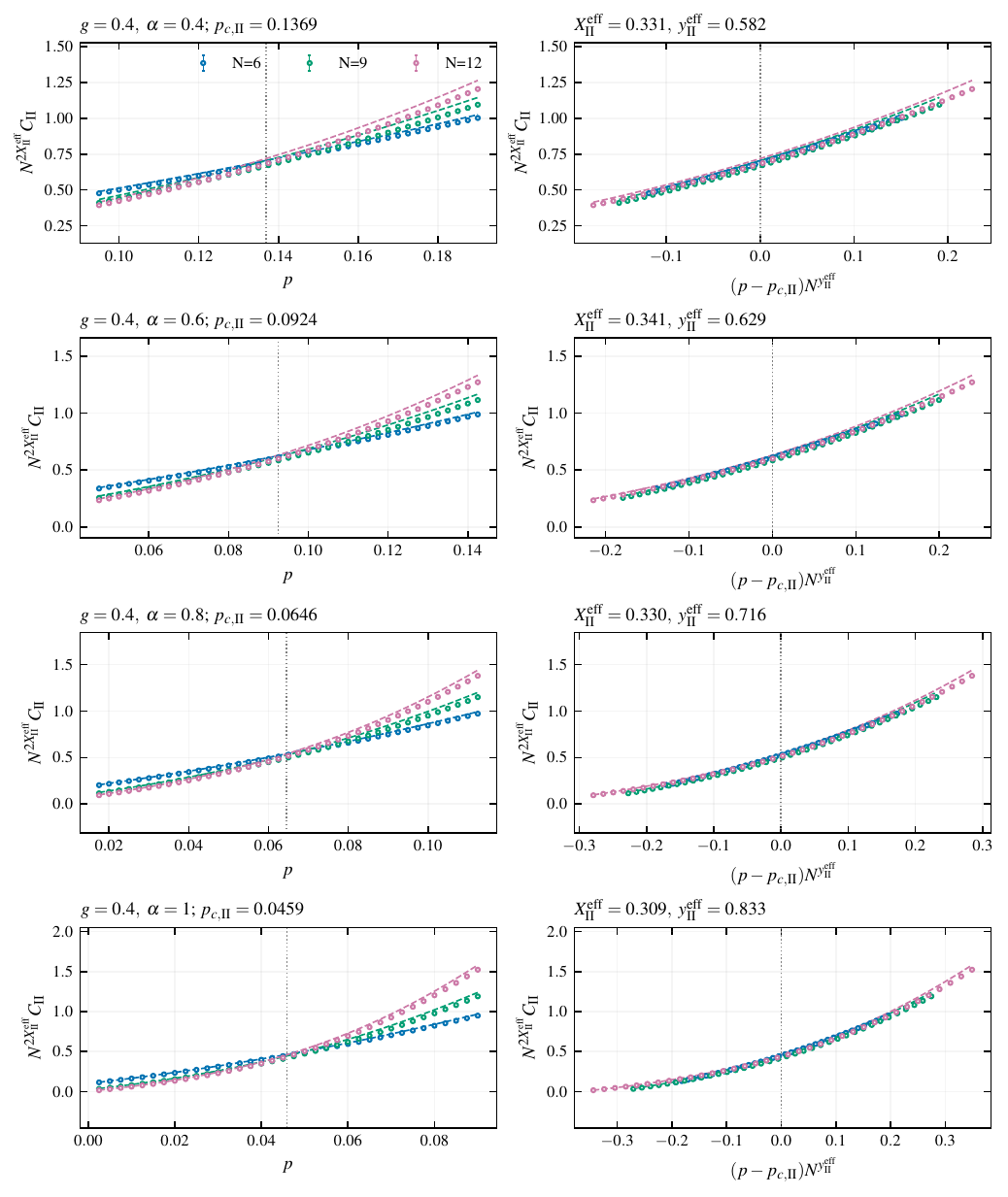}{fig:S-atlas-04-1}{0.4}{0.4,0.6,0.8,1.0}
\boundaryatlas{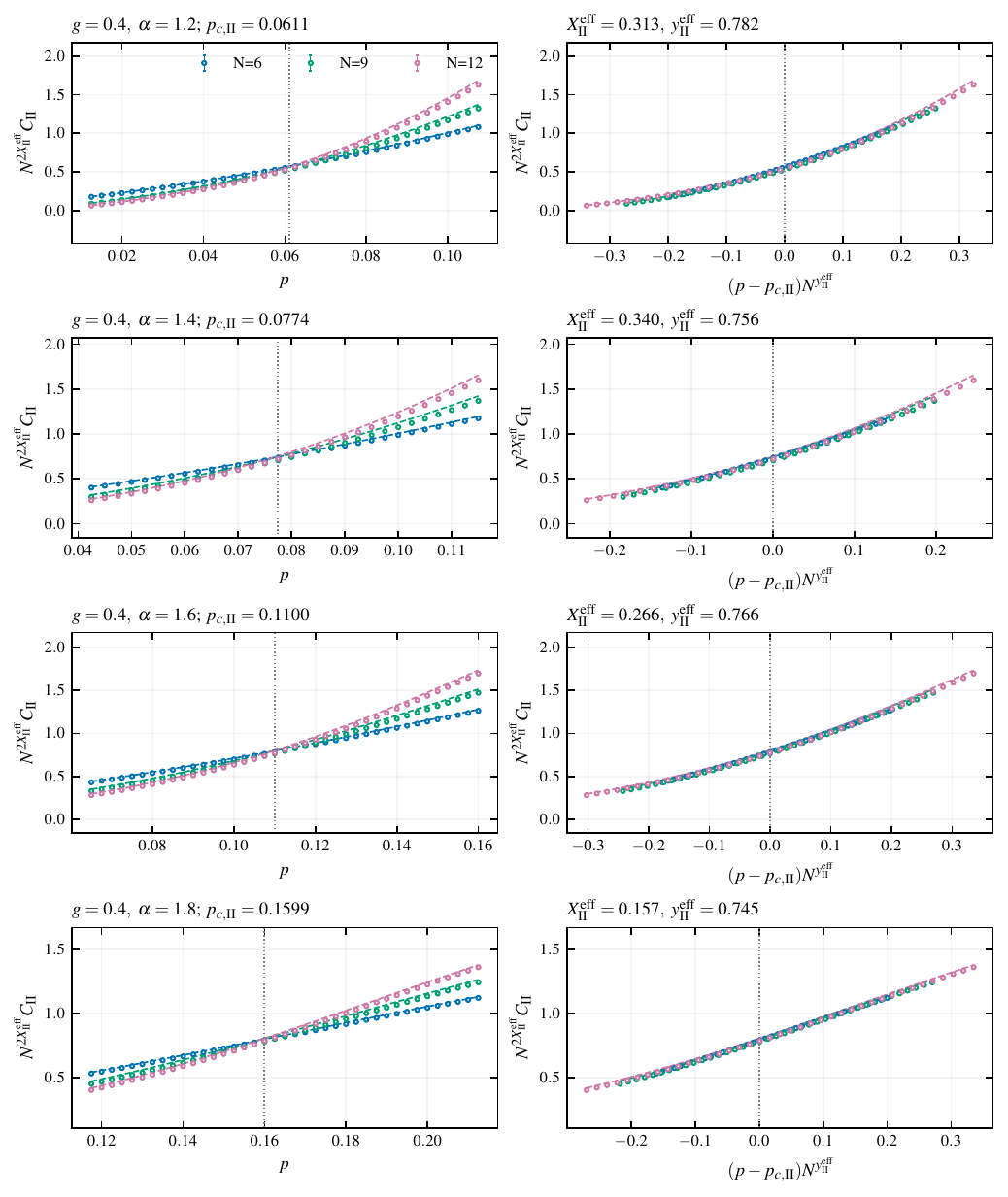}{fig:S-atlas-04-2}{0.4}{1.2,1.4,1.6,1.8}
\boundaryatlas{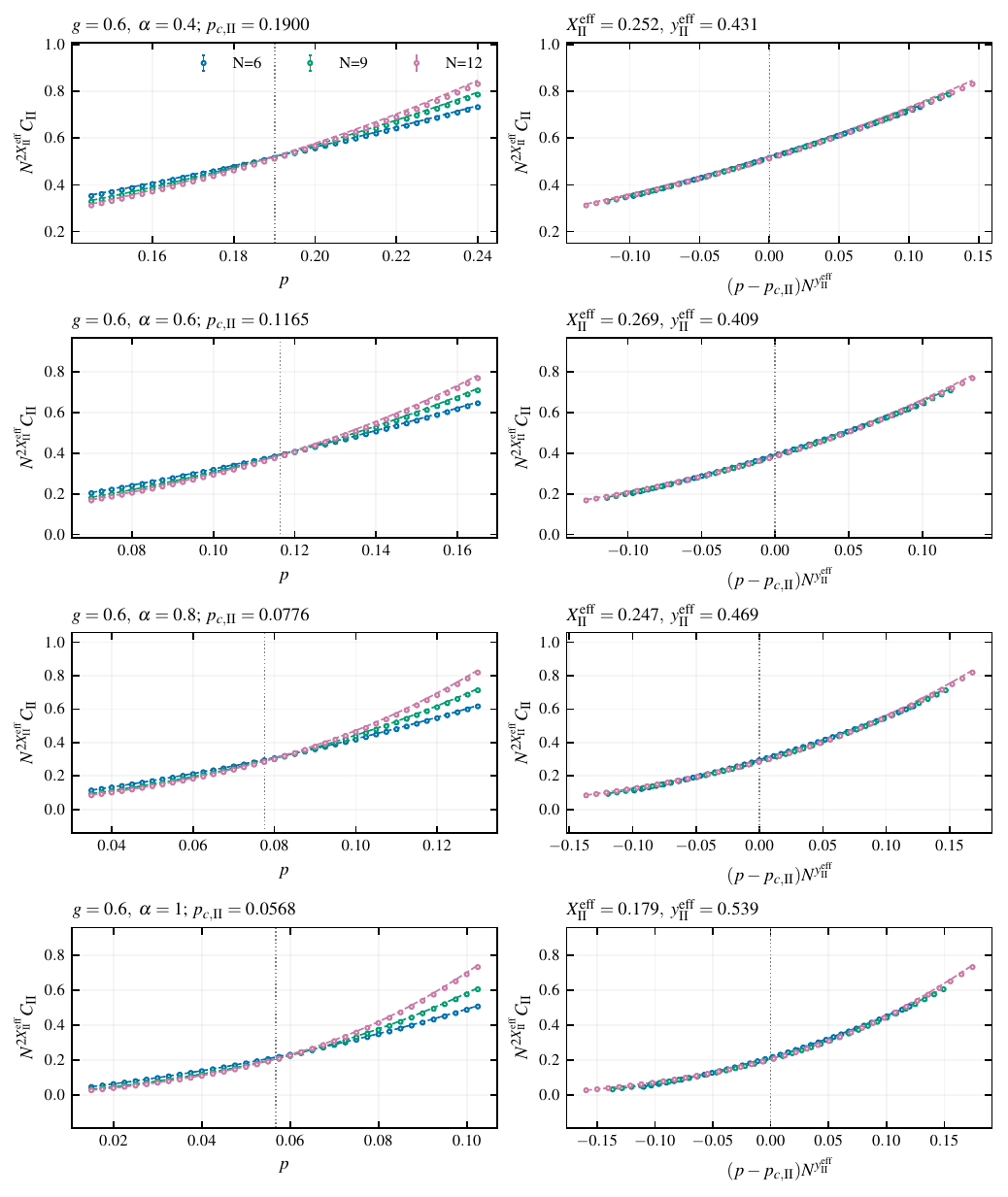}{fig:S-atlas-06-1}{0.6}{0.4,0.6,0.8,1.0}
\boundaryatlas{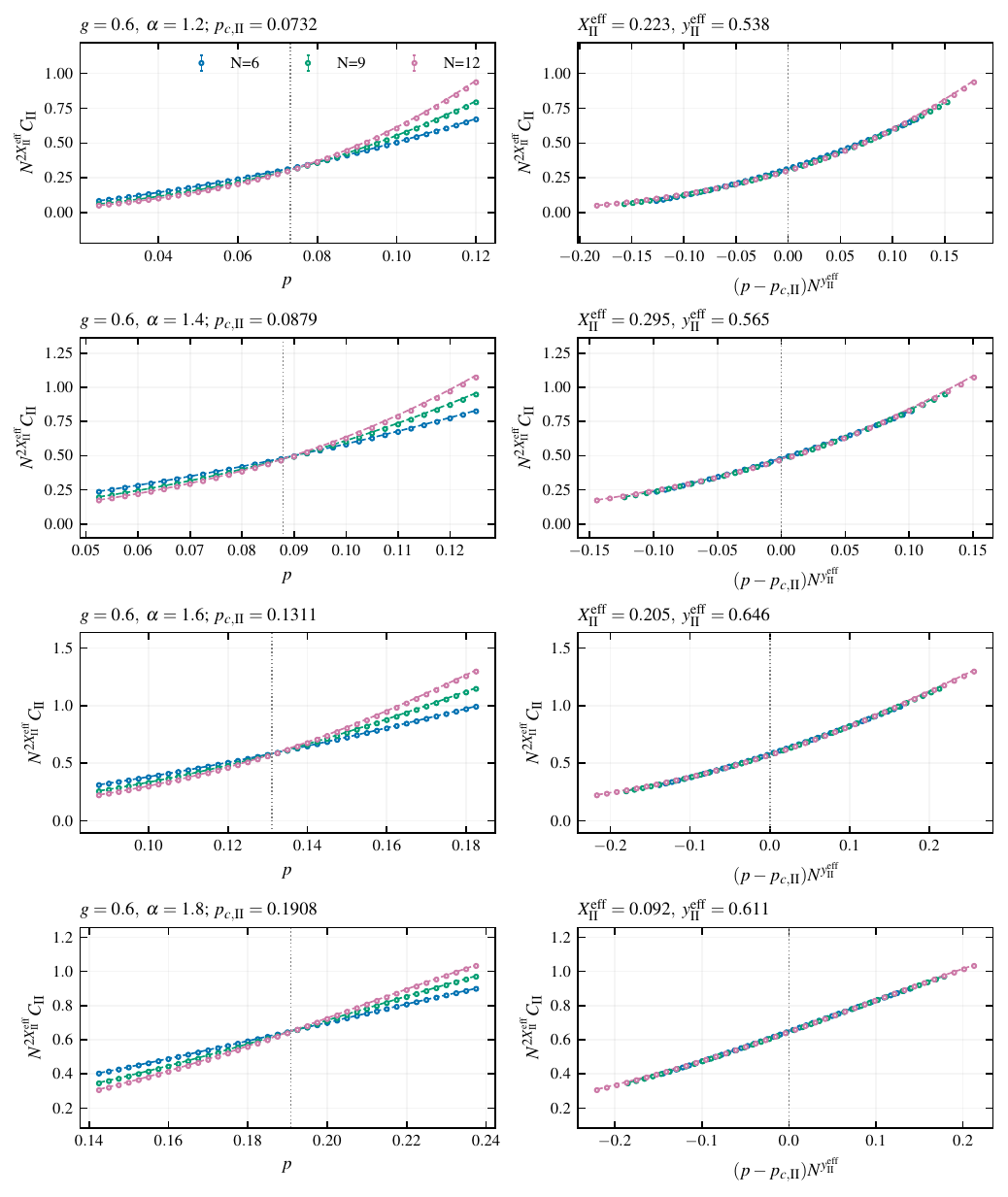}{fig:S-atlas-06-2}{0.6}{1.2,1.4,1.6,1.8}
\boundaryatlas{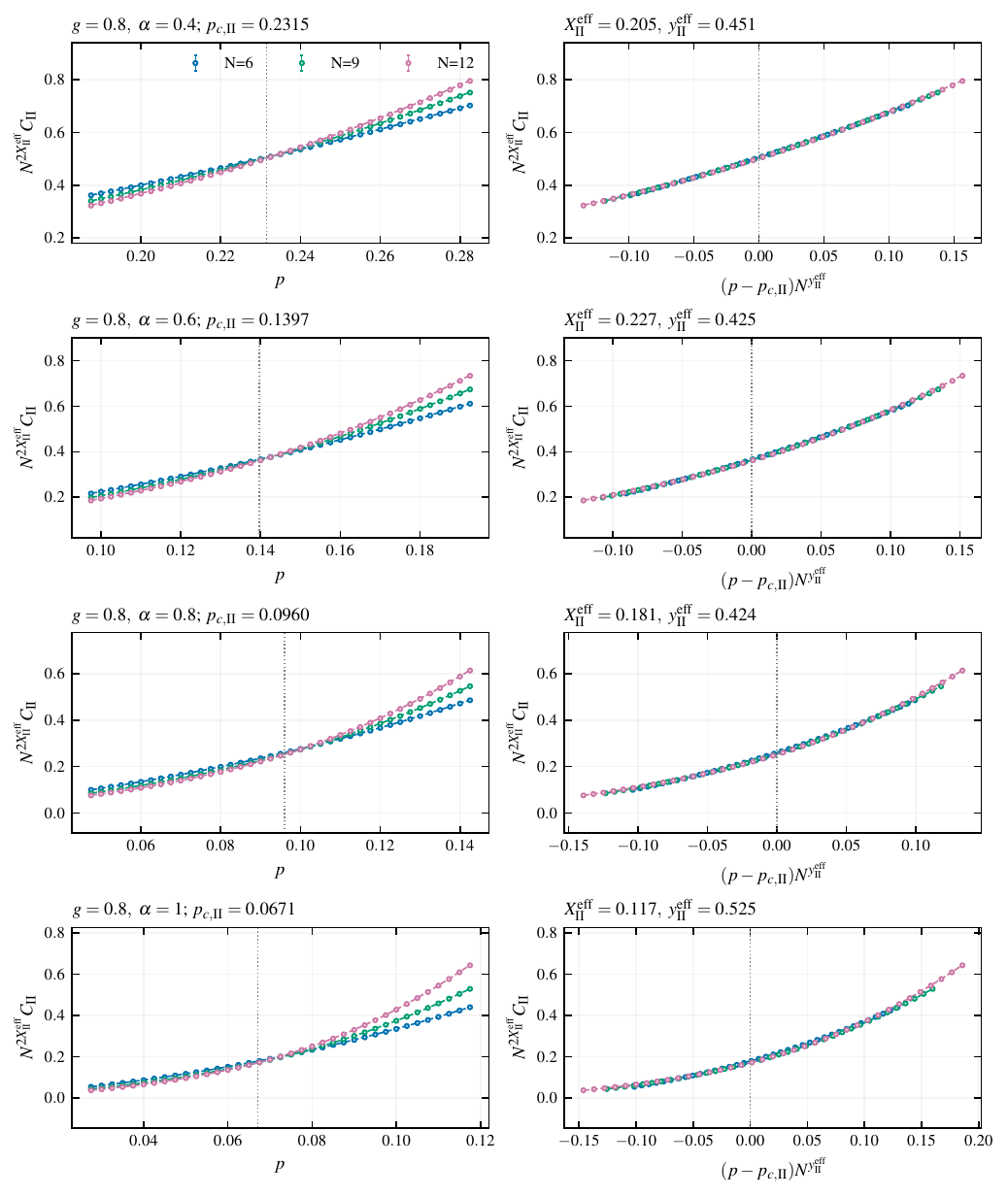}{fig:S-atlas-08-1}{0.8}{0.4,0.6,0.8,1.0}
\boundaryatlas{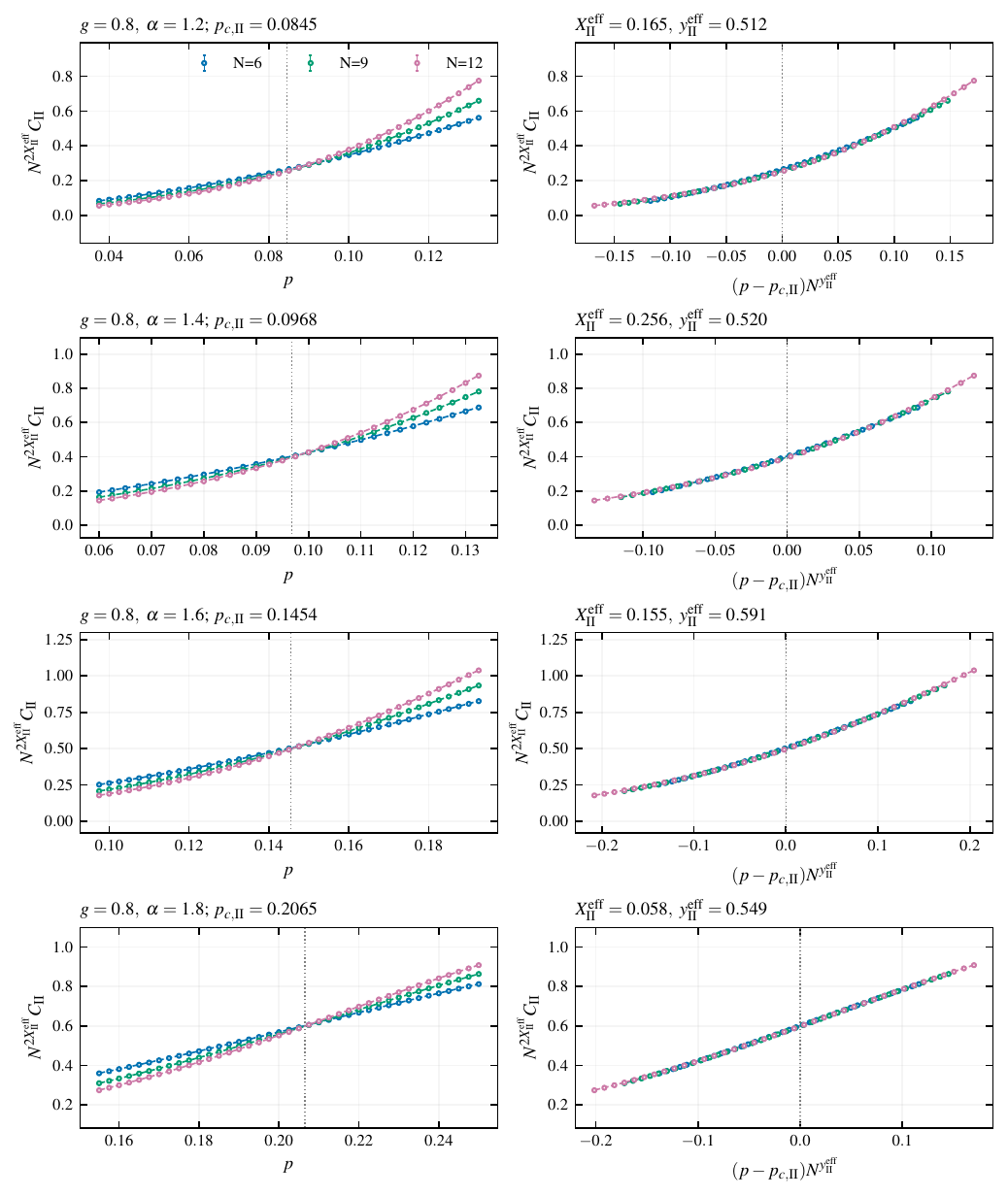}{fig:S-atlas-08-2}{0.8}{1.2,1.4,1.6,1.8}

\newcommand{\fidelityatlas}[3]{%
\begin{figure}[p]
\centering
\includegraphics[width=0.96\textwidth]{figures/#1}
\caption{\label{#2}
$C_F$ finite-size scaling at $g=#3$, using the printed
$(p_{c,F},X_F^{\mathrm{eff}},y_F^{\mathrm{eff}})$
in both columns. Symbols are QPU values, dashed curves are exact
finite-chain references, and bars show statistical standard errors.
Blue, green, and magenta denote $N=6,9,12$.
The guide marks the fitted transition point, which is mapped to zero
after rescaling. Display points can extend beyond the fitting window.}
\end{figure}}

\fidelityatlas{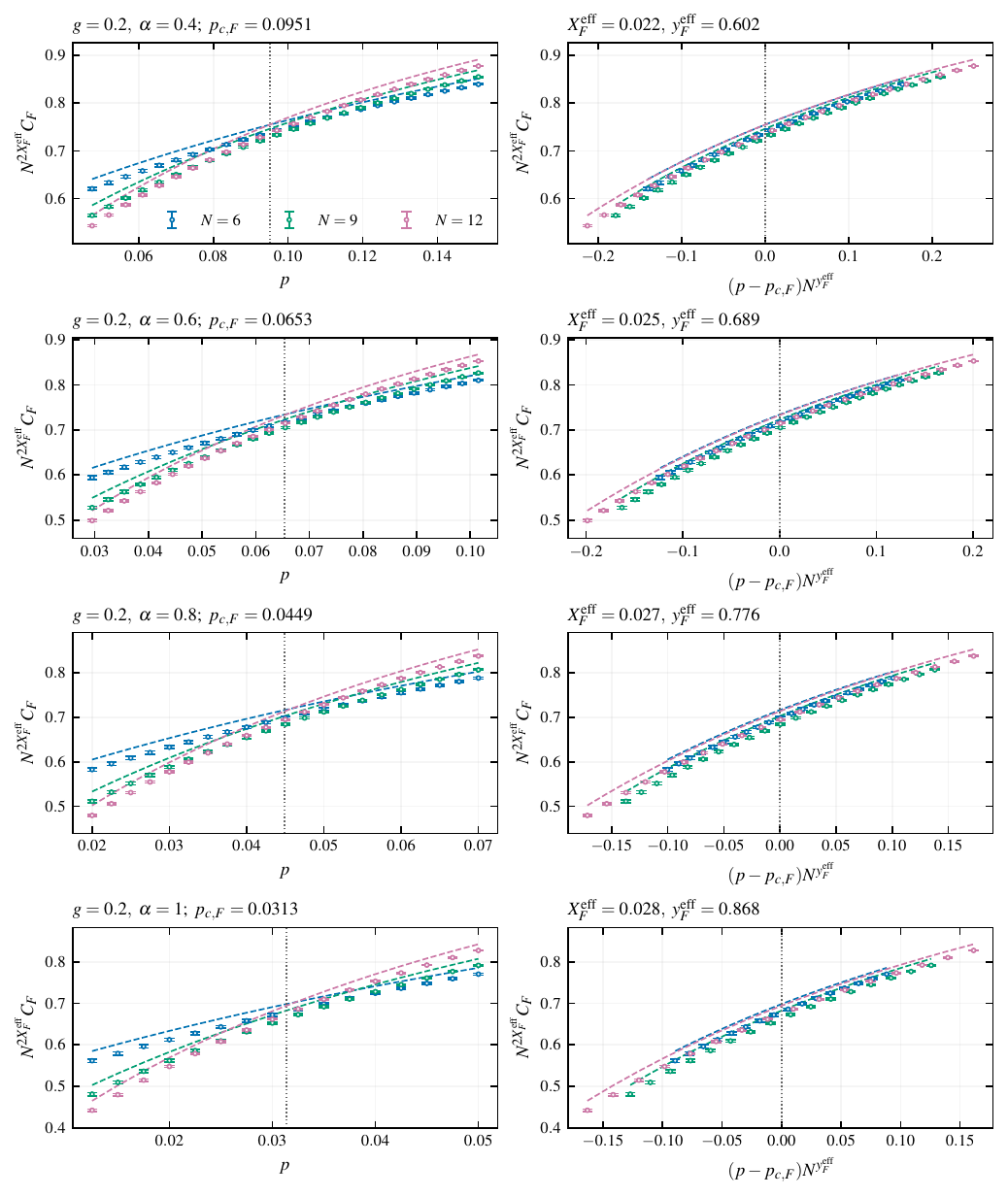}{fig:S-cf-atlas-02-1}{0.2}
\fidelityatlas{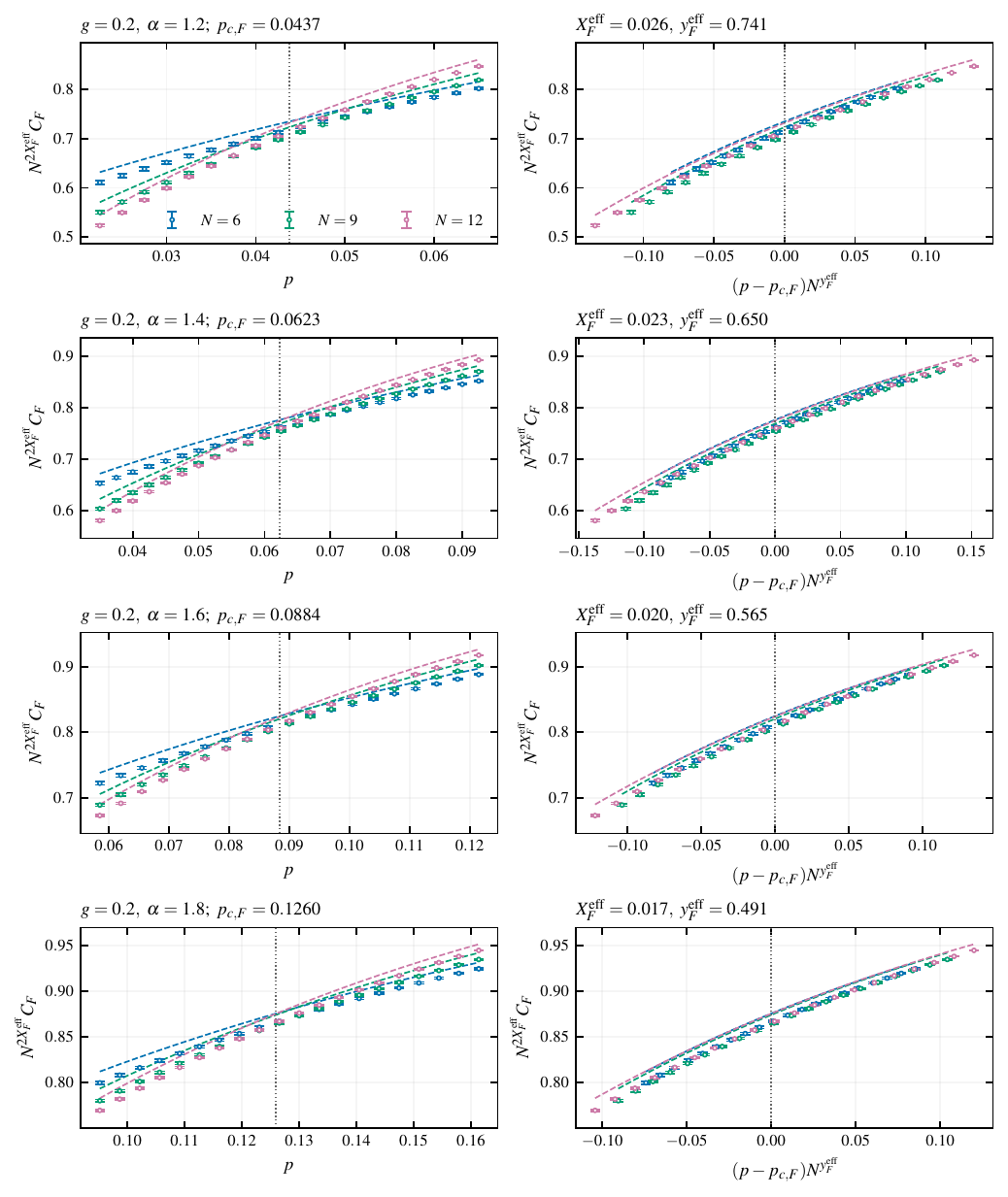}{fig:S-cf-atlas-02-2}{0.2}
\fidelityatlas{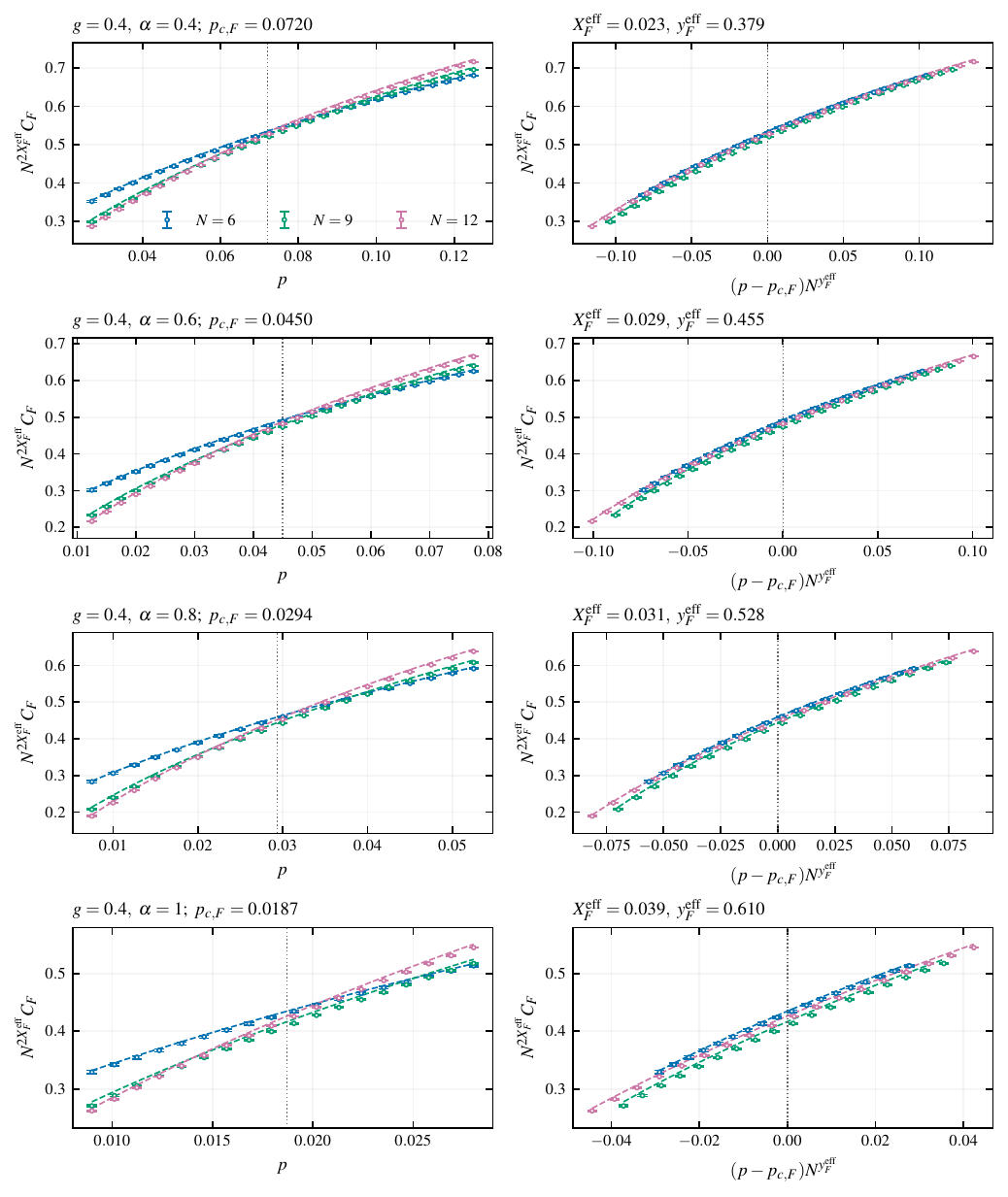}{fig:S-cf-atlas-04-1}{0.4}
\fidelityatlas{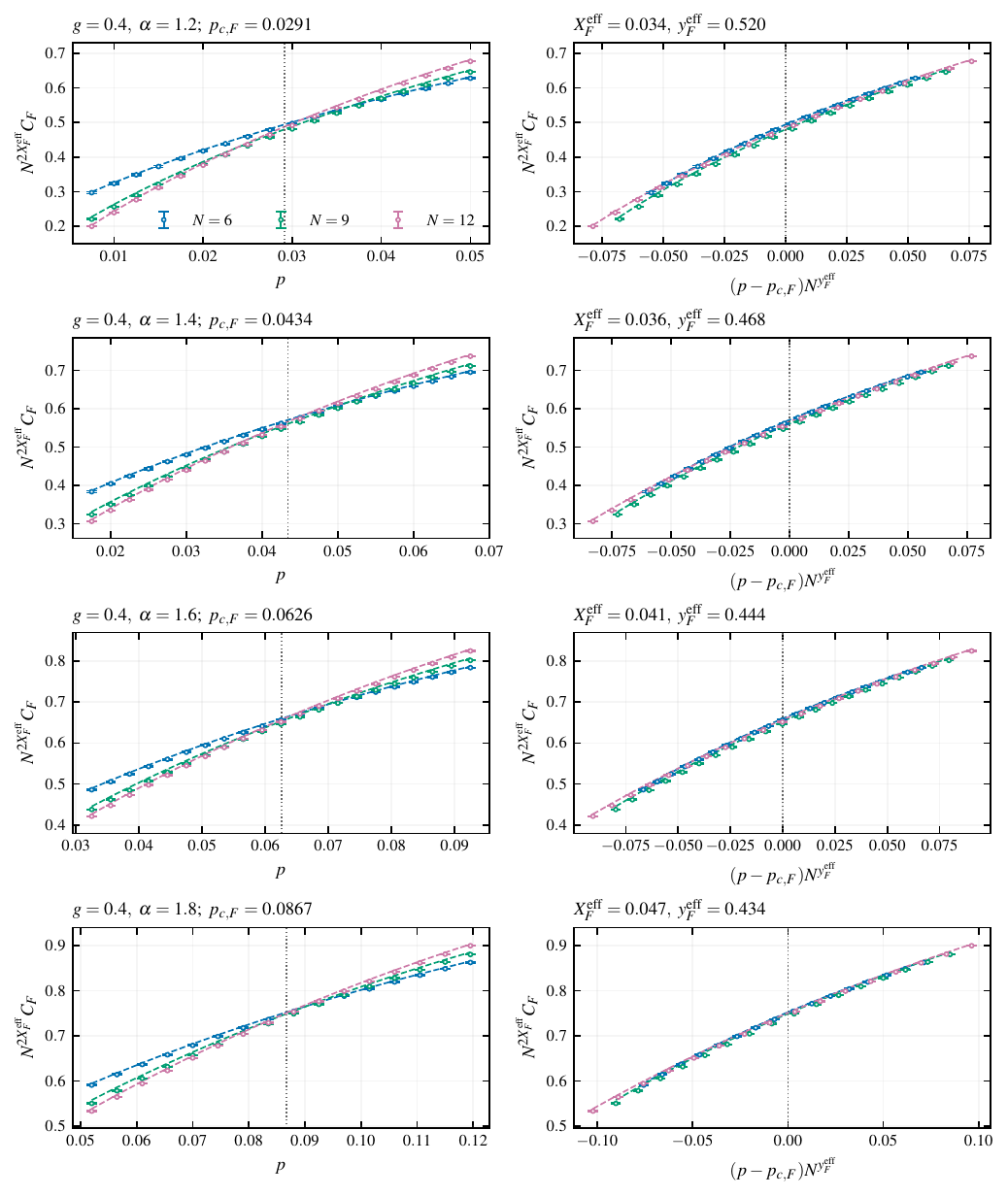}{fig:S-cf-atlas-04-2}{0.4}
\fidelityatlas{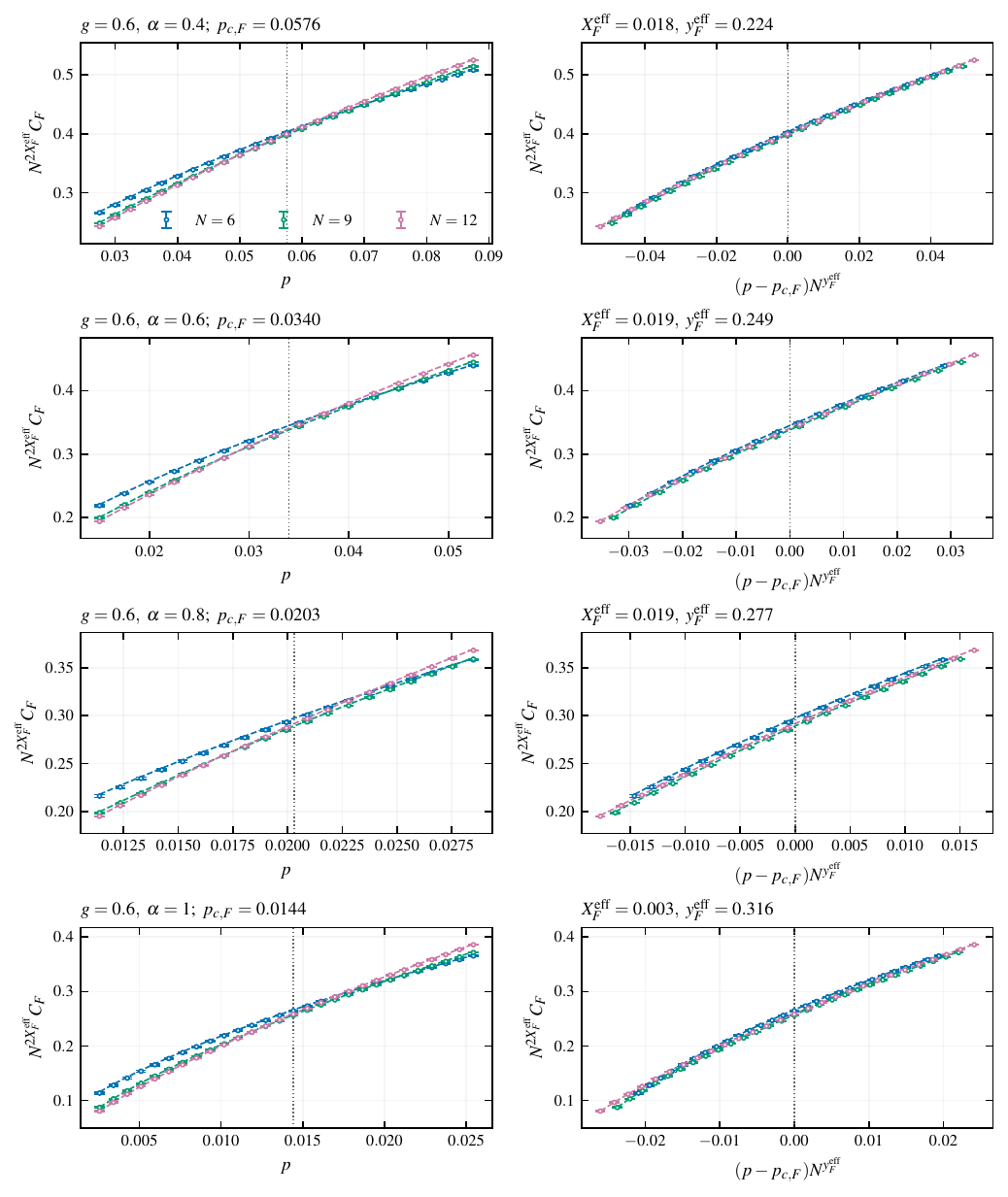}{fig:S-cf-atlas-06-1}{0.6}
\fidelityatlas{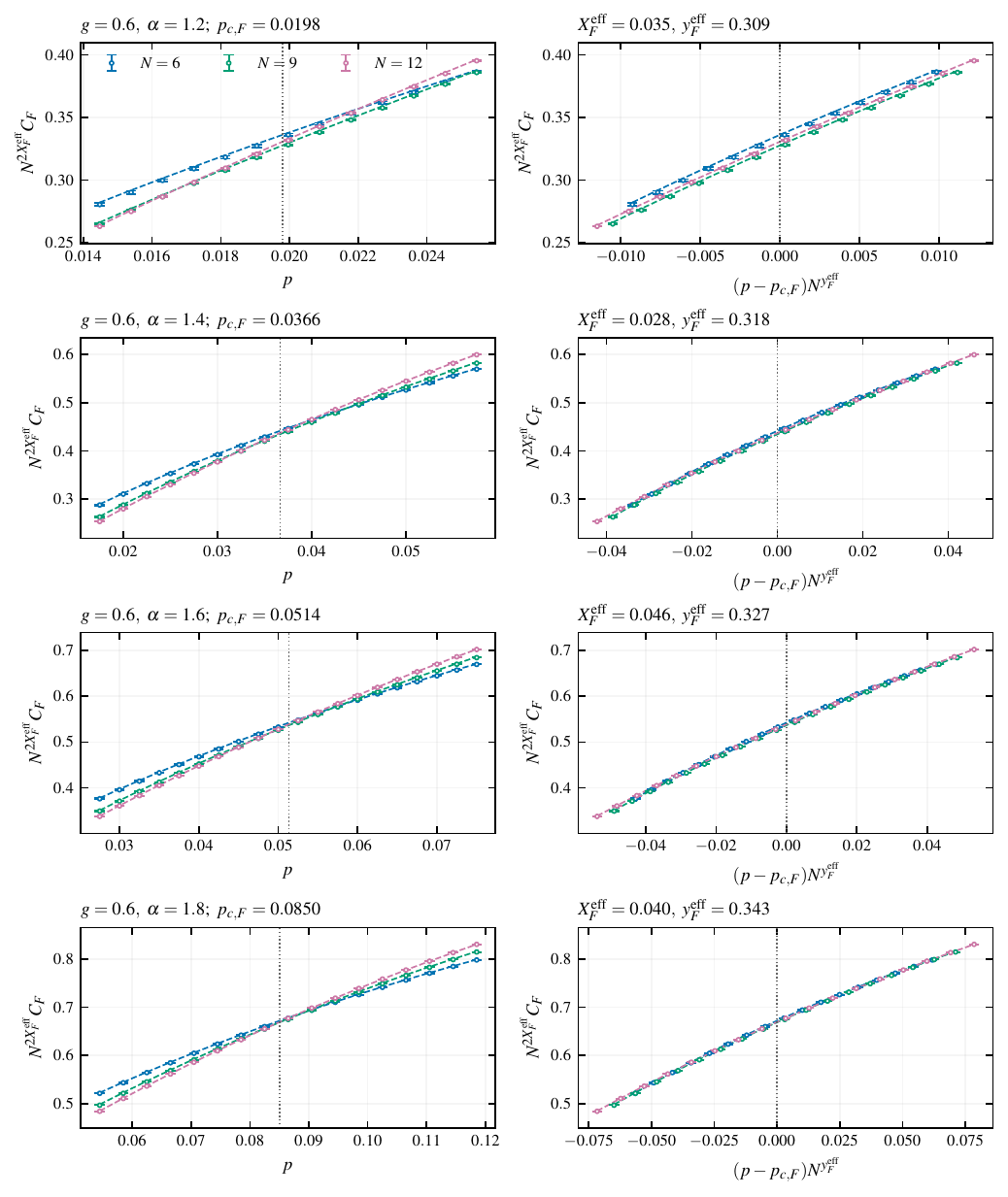}{fig:S-cf-atlas-06-2}{0.6}
\fidelityatlas{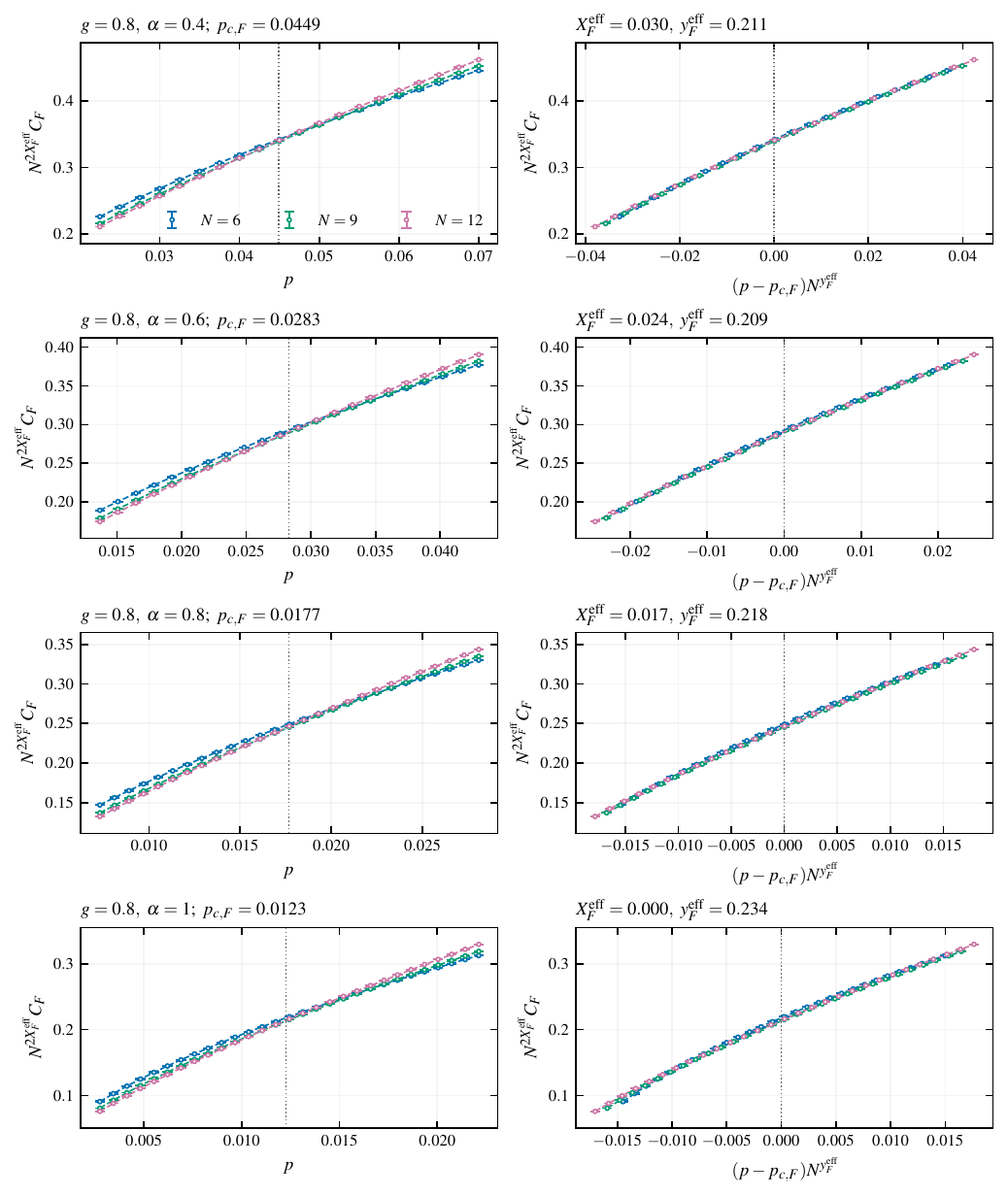}{fig:S-cf-atlas-08-1}{0.8}
\fidelityatlas{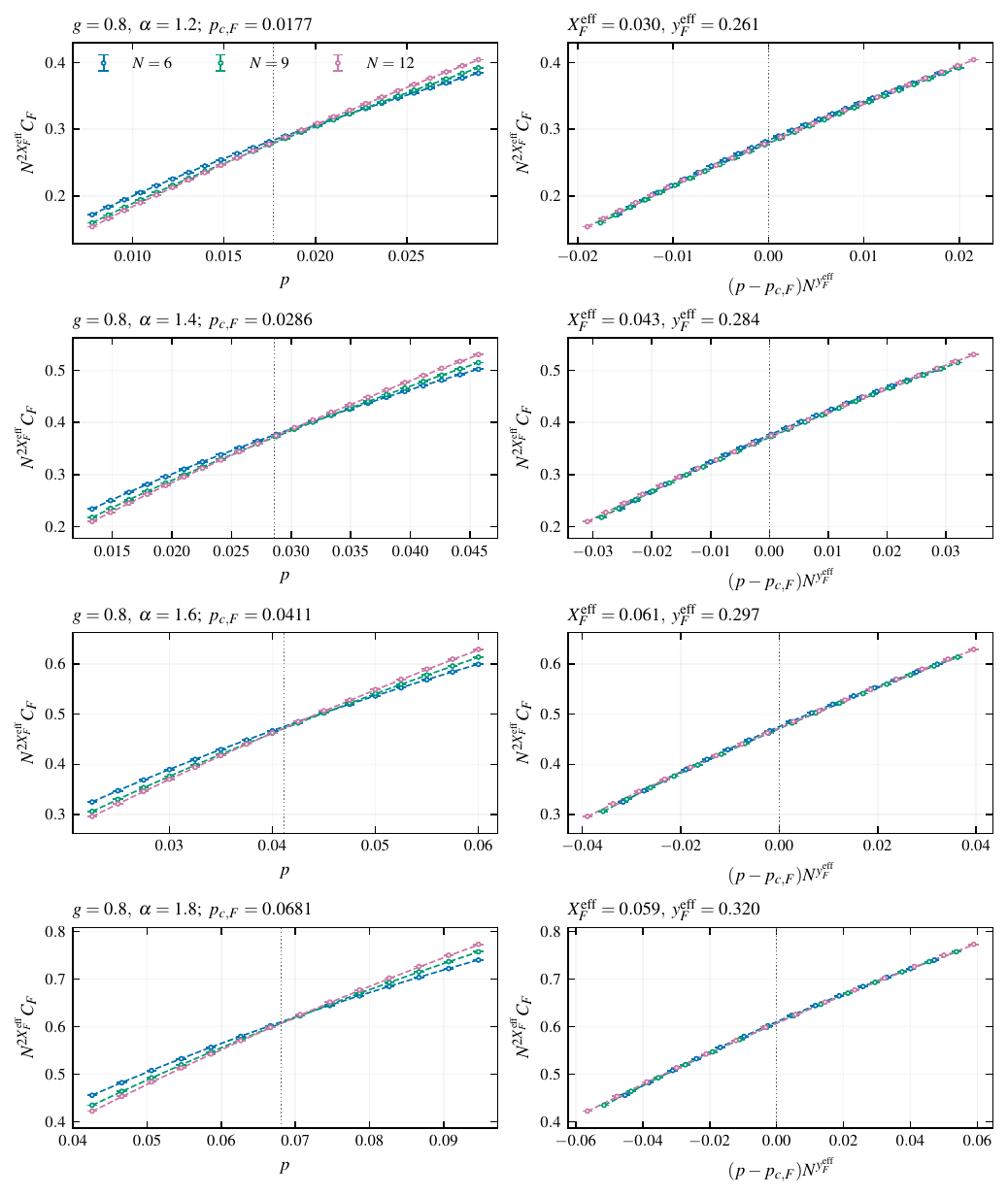}{fig:S-cf-atlas-08-2}{0.8}

\begin{figure}[p]
\centering
\includegraphics[width=0.94\textwidth]{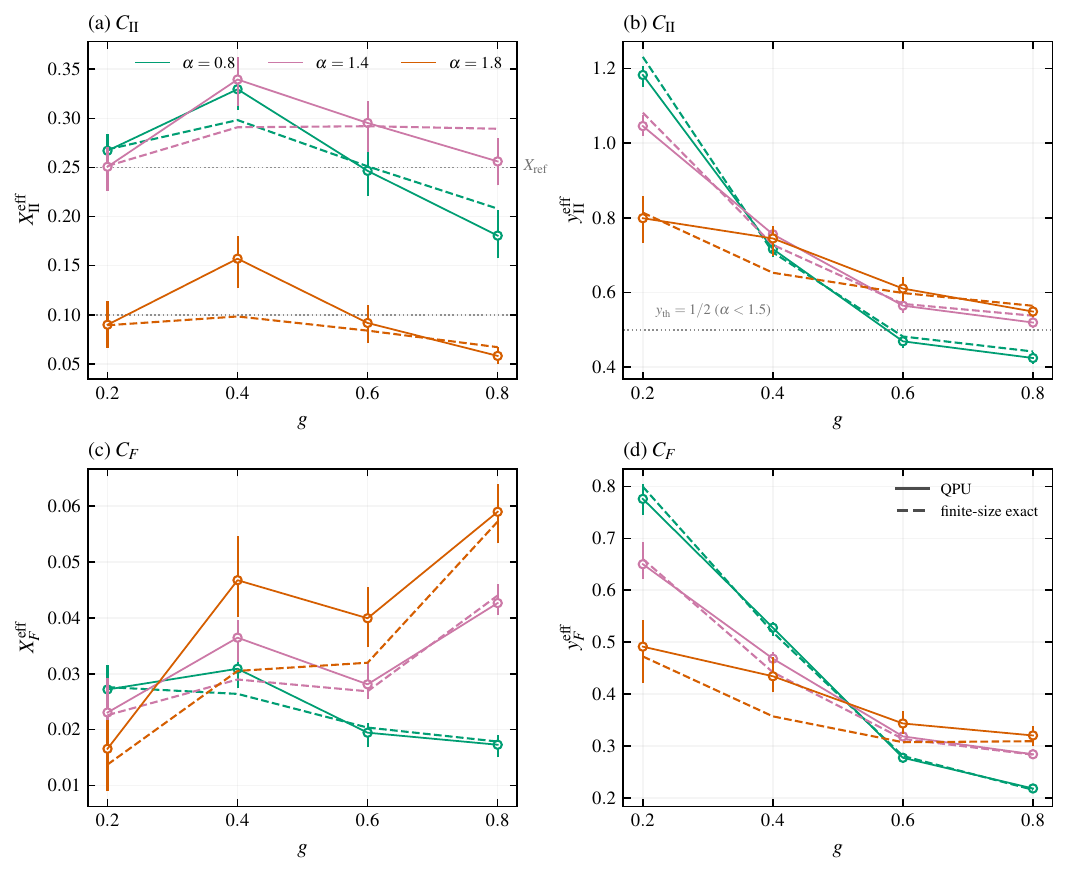}
\caption{\label{fig:S-exponents}
Effective finite-size exponents at $\alpha=0.8,1.4,1.8$.
(a),(b) R\'enyi-2 exponents from regularized three-size fits;
(c),(d) fidelity exponents from {three-size fits with the narrow $X_F$ regularization ($w_F=0.03$)}.
Symbols and solid lines show QPU-derived
exponents, while dashed curves show
exponents obtained from separate fits
to the exact $N=6,9,12$ correlators
using the same fitting windows,
weights, and constraints. In panel (a), thin gray dotted lines mark the
$X_{\rm ref}(\alpha)$ values used for the R\'enyi-2 soft regularization
[Eq.~\eqref{eq:S-regularizer}]: $0.25$ for $\alpha=0.8,1.4$ and
$0.10$ for $\alpha=1.8$. They are not independent fit results.
The gray dotted line in panel (b) marks the asymptotic
$y_{\mathrm{II}}=1/2$ benchmark for $\alpha<1.5$
from Ref.~\cite{fechisin2026powerlaw}; it is not a fit constraint and
is not a prediction for $\alpha=1.8$.
Bars show statistical fit intervals. The corresponding
fitted transition points appear in Fig.~\ref{fig:phase}.}
\end{figure}

\begin{figure}[p]
\centering
\includegraphics[width=0.9\textwidth]{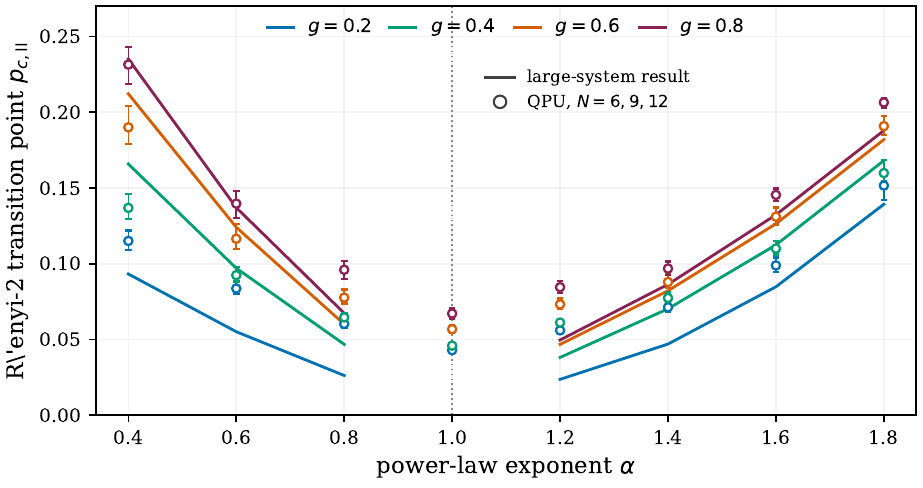}
\caption{\label{fig:S-theory-alpha}
R\'enyi-2 SW-SSB transition point versus the power-law exponent.
{Solid curves connect the large-system R\'enyi-2
transition estimates of Ref.~\cite{fechisin2026powerlaw} at}
$\alpha=0.4,0.6,0.8,1.2,1.4,1.6,1.8$. Open symbols show the QPU
joint-fit results for $N=6,9,12$ from Fig.~\ref{fig:phase}(b), with
statistical fit intervals. Colors denote $g$ in both data sets.
{ Both data sets
show a nonmonotonic trend, with quantitative finite-size shifts.}}
\end{figure}

\begin{figure}[h]
\centering
\includegraphics[width=0.9\textwidth]{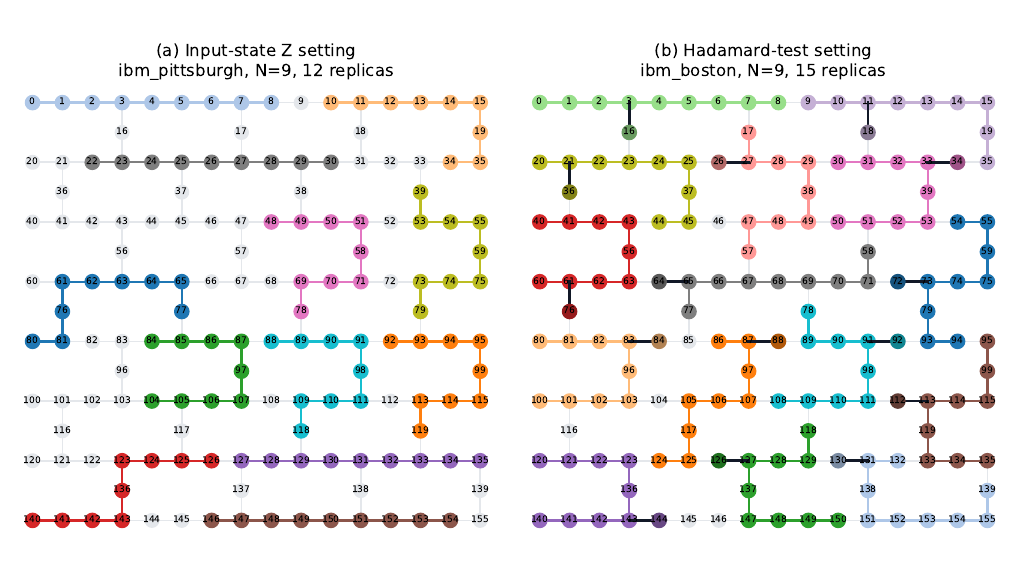}
\caption{\label{fig:S-layout}
{Representative physical layouts for the two measurement
settings, using the actual submitted $N=9$ circuit assignments.
(a) Twelve disjoint MPS preparation chains on \texttt{ibm\_pittsburgh},
using the archived $g=-0.4$ $Z$-basis acquisition that also enters the
Hadamard string analysis. The same type of $Z$-basis setting measures
$C_F$, $C_{\mathrm{II}}$, and $C_{\mathrm I}$ on the positive-$g$ branch,
with the assignments used in those campaigns.
(b) Fifteen disjoint chain-plus-ancilla units on \texttt{ibm\_boston},
used for the $g=-0.8,-0.4$ Hadamard-test string acquisitions.
Each chain has a different color. In (b), the darker node is the
ancilla, and its black link connects to the parity-collection site
$k^\star$. Numbered nodes are physical qubits; colored edges are
used chain couplers. Gray edges show nominal processor connectivity.
Other sizes and campaigns can use different assignments.}
}
\end{figure}

\begin{figure}[p]
\centering
\includegraphics[width=0.97\textwidth,height=0.82\textheight,keepaspectratio]{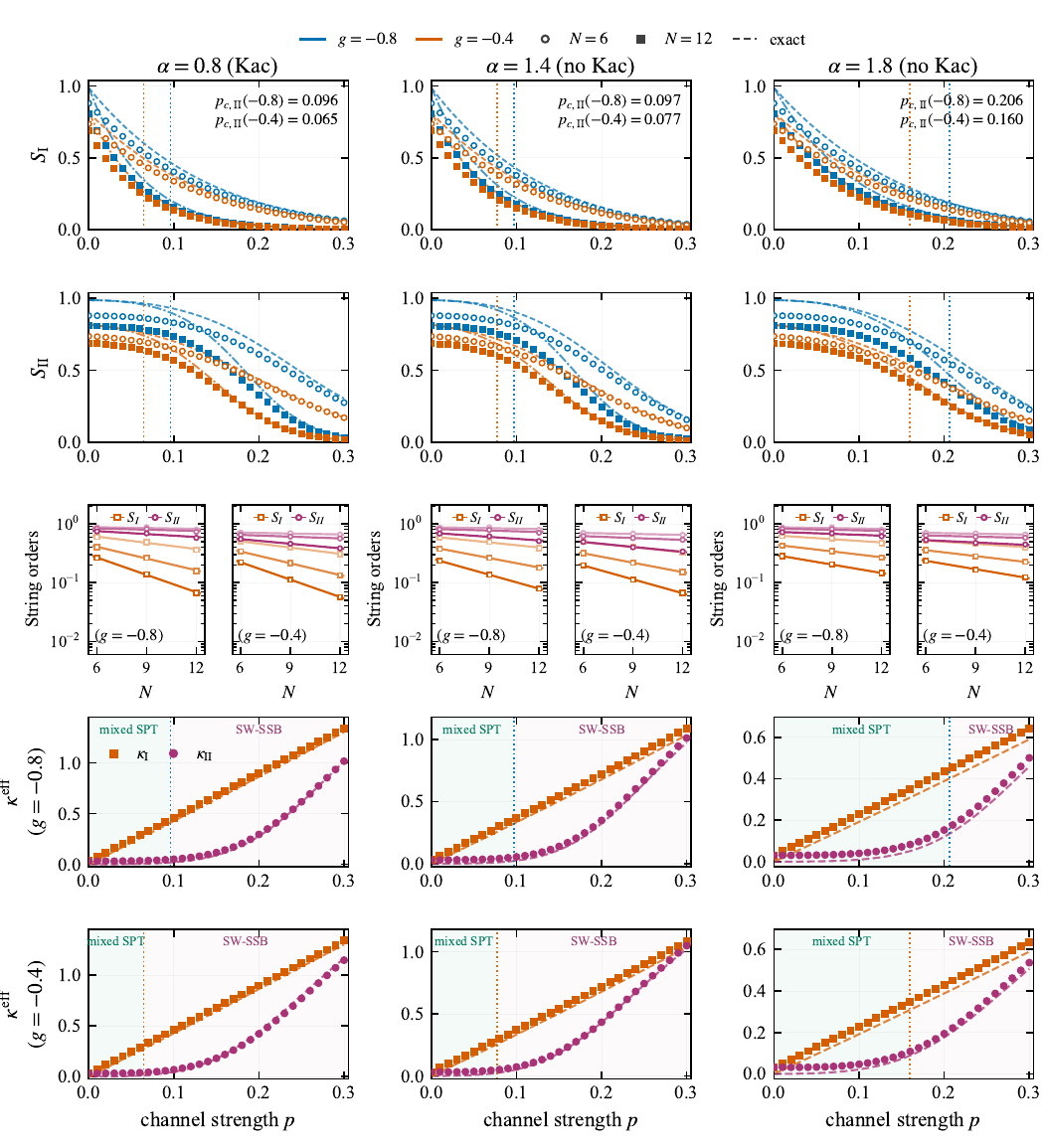}
\caption{\label{fig:S-alpha-strings}
{Mixed-SPT string order measurements at $g=-0.8,-0.4$.
Columns show $\alpha=0.8,1.4,1.8$. The first two rows give
$S_{\mathrm I}$ and $S_{\mathrm{II}}$ for $N=6,12$; blue and orange
denote $g=-0.8$ and $-0.4$. The third row shows the $N=6,9,12$
dependence at $p=0.05,0.10,0.15$ (light to dark), with $g=-0.8$
and $-0.4$ in the left and right subpanels. Squares denote
$S_{\mathrm I}$ and circles $S_{\mathrm{II}}$; solid curves are the
fits of Eq.~\eqref{eq:S-kappa}. The bottom two rows give the resulting
decay rates for $g=-0.8$ and $-0.4$. Dashed curves are exact
finite-chain references. Bars show statistical standard errors,
propagated through the slope fits as described in
Sec.~\ref{sec:S-uncertainty}. Vertical guides mark the independent
$C_{\mathrm{II}}$ transition estimates at the same $|g|$, mapped by
duality~\cite{fechisin2026powerlaw}. The mixed-SPT and SW-SSB labels
refer to this R\'enyi-2 boundary reference.}}
\end{figure}

\clearpage
\end{document}